\documentclass[12pt,aps,a4paper, floats,nofootinbib,amssymb,superscriptaddress]{revtex4}
\pdfoutput=1

\usepackage[sort&compress]{natbib}
\usepackage{epsf,epsfig}
\usepackage{amsmath}
\usepackage{hyperref}
\usepackage{subfigure}
\usepackage{graphicx}
\usepackage{color}
\usepackage{mathrsfs}

\newcommand{\be}{\begin{equation}}
\newcommand{\ee}{\end{equation}}
\newcommand{\bea}{\begin{eqnarray}}
\newcommand{\eea}{\end{eqnarray}}

\newcommand{\ba}{\begin{array}}
\newcommand{\ea}{\end{array}}
\newcommand{\bi}{\begin{itemize}}
\newcommand{\ei}{\end{itemize}}

\newcommand{\wplus}{W^{+}}
\newcommand{\wminus}{W^{-}}

\newcommand{\myspace}{\vspace{0.2cm} \\}

\newcommand{\ih}{-\frac{i}{2}}

\makeatletter

\begin{document}


\title{\vspace*{1.in} 
 Introducing tools to test Higgs interactions via $WW$ scattering I:
  one-loop calculations and renormalization in the HEFT
\vspace*{0.7cm}
}

\author{I\~nigo  Asi\'ain}\email{iasiain@icc.ub.edu}\affiliation{Departament de F\'isica Qu\`antica i Astrof\'isica\,,
Institut de Ci\`encies del Cosmos (ICCUB), \\
Universitat de Barcelona, Mart\'i Franqu\`es 1, 08028 Barcelona, Spain}
\author{Dom\`enec Espriu}\email{espriu@icc.ub.edu}
\affiliation{Departament de F\'isica Qu\`antica i Astrof\'isica\,,
Institut de Ci\`encies del Cosmos (ICCUB), \\
Universitat de Barcelona, Mart\'i Franqu\`es 1, 08028 Barcelona, Spain}
\author{Federico Mescia}\email{mescia@ub.edu}\affiliation{Departament de F\'isica Qu\`antica i Astrof\'isica\,,
Institut de Ci\`encies del Cosmos (ICCUB), \\
Universitat de Barcelona, Mart\'i Franqu\`es 1, 08028 Barcelona, Spain}
\thispagestyle{empty}

\begin{abstract}
\vspace*{1.0cm}
\noindent
Effective field theories are useful tools to search for physics beyond the Standard Model (SM). 
However, effective theories can lead to nonunitary behavior
with fast growing amplitudes. This unphysical behavior may induce a too large sensitivity to
SM deviations, making necessary a unitarization of the amplitudes prior to a comparison with
experiment. In the present work, we focus on all the processes entering two-Higgs production
via longitudinal $WW$ scattering.
We perform a one-loop calculation in the HEFT framework of all relevant processes,  
determining the necessary counterterms in the on-shell scheme, 
and we study how the full inclusion of the gauge degrees of freedom modifies the previously computed
masses and widths of the dynamical resonances arising from the unitarization process in the vector-isovector channel.
Altogether, we are able to provide the technical tools that are needed to study the low-energy
couplings in the Higgs effective theory under the requirements of unitarity and causality.   
\end{abstract}

\maketitle

\newpage
\thispagestyle{empty}
\tableofcontents

\thispagestyle{empty}
\newpage
 \setcounter{page}{0}


\section{Introduction}
\label{sec:intro}

Since the discovery in 2012 of a light scalar by ATLAS \cite{ATLAS:2012yve} and CMS~\cite{CMS:2012qbp}, so far
compatible with the
Standard Model (SM) Higgs, a lot of questions have arisen regarding the origin of such a scalar and hence the properties 
of the electroweak symmetry breaking sector (EWSBS)~\cite{Giudice:2007fh, Alonso:2012px, Pich:2012dv, 
Buchalla:2013rka, Contino:2013kra, Buchalla:2016bse, Pich:2018ltt}. 
\myspace
To explore the  nature of EWSBS beyond the SM (BSM), the scattering of longitudinally polarized electroweak gauge bosons is 
one of the most sensitive channels. 
The appearance of heavy resonances in the scattering of longitudinally polarized gauge bosons, for example, will be a 
clear indication of the existence of a strong dynamics behind EWSB~\cite{Espriu:2012ih,  Arnan:2015csa, Pich:2015kwa, 
Pich:2016lew,  Pich:2020xzo}.\myspace 
The main properties of these resonances can be studied using effective theory treatment together  with partial wave 
analysis and unitarization techniques~\cite{Espriu:2012ih, Espriu:2013fia, Espriu:2014jya, Delgado:2015kxa, 
Corbett:2015lfa, Delgado:2017cls, Garcia-Garcia:2019oig}. 
Over the years, the use of inverse amplitude method (IAM) to build unitarity amplitudes has been successfully applied 
to explain the resonances in the pion-pion scattering~\cite{Salas-Bernardez:2020hua, Dobado:2001rv, Guerrero:1998ei, 
Oller:1998hw, Oller:1997ng, Dobado:1996ps, Dobado:1989qm, Truong:1988zp}. 
The IAM allows us to predict mass and width of dynamically generated resonances from  
the unitarized amplitudes of the low-energy effective theory.  In turn, this also allow us to set  bounds on the  couplings 
of the underlying effective theories.\myspace   
Among the open questions of EWSBS is the nature of the Higgs potential. Even if one assumes that the Higgs-like
scalar found is truly elementary, are its self-interactions the ones predicted by the textbook SM?\myspace
Indeed, this is one of the main purposes of future machines such as the future ILC linear collider in
Japan or the planned FCC ${e^+e^-}$ at CERN~\cite{EuropeanStrategyGroup:2020pow}. In all these cases, setting bounds 
on effective couplings needs
a {\em bona fide} and fair comparison that requires using unitarized amplitudes when departures from the SM
values could be potentially large.
The reason is that deviations of the couplings in the effective theory from their SM values lead to rapidly
increasing cross sections and this may artificially enhance the sensitivity to the said couplings. \myspace
The purpose of this article is to provide some tools that would make this comparison possible. More specifically, 
we compute the renormalization counterterms at one loop that are required to calculate the processes 
$W_L W_L \to W_L W_L$, $ W_L W_L \to h h$
and $hh\to hh$. All these amplitudes enter the  unitarization  of the $I,\, J=0,0$ channel.
Moreover, we also provide for these $2\to 2$ processes the corresponding renormalized amplitudes. Here, we calculate 
the full ${\cal O}(g^2)$ contributions of these processes.
This paper thus completes and extends our previous work in  Refs.~\cite{Espriu:2013fia, Espriu:2012ih, Espriu:2014jya}, 
where the analysis was carried out in the limit of no gauge interactions, $g=0$; only the longitudinal parts of the EW bosons, i.e 
Goldstone bosons (GBs), were taken into account inside the loops. In the present work, as mentioned, we relax that 
approximation and allow transverse modes to propagate in the process, improving a weak point of the previous 
unitarization studies because along with the assumption $g=0$ in~\cite{Espriu:2013fia, Espriu:2012ih, Espriu:2014jya}, 
the authors consistently set $M_W=0$ for the real part of the loop calculation. 
With respect to previous works, we also compute the processes involving double Higgs production.\myspace
The derivation is made in the framework of the Higgs effective field theory (HEFT)~\cite{Giudice:2007fh, Alonso:2012px, 
Pich:2012dv, Buchalla:2013rka, Contino:2013kra, Buchalla:2016bse, Pich:2018ltt, Pich:2015kwa}, where the global symmetries are
nonlinearly realized and the complex doublet structure of the SM is not assumed \textit{a priori} (see e.g. the discussions
in \cite{Dobado:2019fxe}). 
The calculation of the $2\to 2$ physical amplitudes needed at the one-loop level beyond the SM
is relatively involved. Thus, it is useful to take some shortcuts in order to have more manageable expressions.
The real part of the amplitudes will be computed using the equivalence 
theorem~\cite{He:1993qa,  Grosse-Knetter:1994lkr, Dobado:1994vr, Dobado:1993dg, Chanowitz:1985hj, Gounaris:1986cr, 
Lee:1977eg, Cornwall:1974km} (ET), where  the longitudinal components of $W$ in the external states are substituted 
by their Goldstone bosons (GBs). This approach is fully consistent in order to study cross-section of longitudinal 
polarized $W$ at energies much larger of the EW scale. 
The imaginary part of our amplitude is exactly obtained  via the optical theorem, where physical $W$ are present 
in the external legs.\myspace
In this work,  we make no assumption about the UV strong dynamics, In our model-independent study, the effects of 
the high-energy theory in the low energy regime are encoded
in the so called chiral parameters. When these parameters do not have a correspondence in the SM, their presence spoils
the unitarity of the amplitudes leading them, after unitarization, to exhibit resonances, i.e., bound states 
presumably resulting from the underlying strong dynamics.\myspace
For the  purposes of this study, the custodial symmetry is assumed to remain exact and the soft breaking of the global
symmetry $SU(2)_L \times SU(2)_R$ induced by the gauging of the hypercharge group will be neglected. We believe this
approximation to be well justified by the experimental results of the $\rho$ parameter. In this limit,
the electromagnetism  is removed from the fundamental interactions and the gauge bosons transform exactly as a triplet
under the vector (or custodial) subgroup after the global symmetry breaking pattern $SU(2)_L \times SU(2)_R \rightarrow SU(2)_V$,
where $V$ stands for $L=R$. The absence of electromagnetic interactions moves the pole of $Z$ to the very same position of
that of the $W$, making the $\rho$ parameter exactly equal to one at every order in perturbation theory.\myspace
As mentioned above, a full derivation of the one-loop counterterms in the HEFT, with a dynamical Higgs, is a necessary 
step in the process. A previous calculation of all the required counterterms does exist in the 
literature~\cite{Buchalla:2020kdh, Buchalla:2017jlu}. However, their expressions are not easily 
translated to the calculation  of physical processes, we are interested in.  For one thing, the renormalization scheme   
is not the widely-used on-shell scheme to which we adhere. On the other hand in \cite{Buchalla:2020kdh, Buchalla:2017jlu} 
extensive use is made of the equations of motion and field redefinitions, including some mixing of operators with different chiral 
dimensions. All this makes their results difficult or impossible to translate to a $S$-matrix calculation. Recently, an independent 
diagrammatic calculation was published \citep{Herrero:2021iqt}, where a large set of counterterms are
derived off shell (but only those needed for elastic vector boson scattering). We will review below our agreement with these preexisting
results. It is worthwhile emphasizing that our approach is purely diagrammatic and inspired by the practical requirements 
needed when $S$-matrix elements are to be computed.\myspace
In the interest of practicality a number of simplifications have been made. They do not impact in any significant
way  on the validity or relevance of the results. Let us list them here for the sake of clarity:
(a) The equivalence theorem has been used to compute the real part of the one-loop correction. This does not restrict
in any way the ability to obtain all the appropriate counterterms and it is an efficient way of arranging
the calculation. This approximation also bypasses some subtleties related to crossing that
will be pointed out below. (b) The equations of motion are systematically used: our results are relevant for on-shell
processes and we actually have nothing to say for off-shell Green functions. Unlike the previous one, this approximation does reduce the number of contributing effective operators respect to the full list that is provided in the references \cite{Gavela:2014uta} and \citep{Herrero:2021iqt}, many of which are redundant when the equations of motion
are used. (c) We work within the HEFT under the approximation of considering
only custodially symmetric operators. This reduces even further the number of required
operators and also implies the so-called isospin limit where $M_W = M_Z$. This approximation is, of course, numerically 
irrelevant for physics in the TeV region. But, in fact, there is a deeper reason that makes this approximation convenient: 
we set $g^\prime=0$ and neglect accordingly electromagnetism because it would not be possible to use the usual isospin 
decomposition otherwise and, in addition, long range interactions
are not easily amenable to unitarization techniques. Obviously, electromagnetism should not be involved in any strong dynamics that
may be present in the EWSBS. (d) Accordingly custodially breaking
operators are not included, as previously indicated; it would be inconsistent to include these and leave out
the main source of weak isospin breaking in the SM. (e) The calculation is made in the Landau gauge, which simplifies somewhat
the counterterm structure.

\section{The effective Lagrangian}\label{sec:lagrangian}
The electroweak chiral Lagrangian is a nonlinear gauged effective field theory mimicking chiral perturbation
theory, used to investigate low-energy QCD, in the electroweak sector \cite{Dobado:1990zh}. It has been used
intensively in the context of effective field theories since the early days of LEP 
\cite{Altarelli:1993sz, Peskin:1991sw, Dobado:1990zh} in order to put to the test extensions
of the SM. In the case of the electroweak sector, it only assumes the local and global properties known to hold at low energies,
and makes no specific commitment to the underlying physics. The addition of the Higgs scalar makes this model a Higgs effective
field theory (HEFT). \myspace
This HEFT contains as dynamical fields, the EW gauge bosons $W^{\pm},Z,\gamma$; their associated Goldstone
partners $\omega^a=\omega^{\pm},z$; and a light Higgs $h$ (the latter could or could not be a Goldstone boson).
In the HEFT the Goldstones resulting from the electroweak breaking are described by a unitary
matrix $U$ that takes values in the coset $SU(2)_L \times SU(2)_R/SU(2)_V$ and the Higgs is a $SU(2)$ singlet.
This fact is in contrast to the textbook SM, where the Higgs is part
of a complex doublet and transforms alongside the GBs. Effective theories describing the Higgs as a part of a $SU(2)$
doublet $\Phi$ are termed Standard Model effective field theories (SMEFT).\myspace
The HEFT is fairly general and its form is largely independent of the details
of the EWSBS because it is based only on the symmetry properties and the fact that only the light degrees
of freedom are retained. The latter depends on the symmetry breaking pattern $G\to H$ and the way the electroweak
gauge group $G_{EW}$ is embedded in $G$. In this type of effective theories, the Higgs may or may not  be a Goldstone boson.
Having light states other than the Higgs (such as e.g. additional Goldstone bosons) could
be worrisome from a phenomenological point of view because it would be very difficult or impossible to find
mechanisms that would make them so massive to be able to escape detection. We should then exclude
such a possibility from the effective theory.  \myspace
When can a particular HEFT  be written in the form of a SMEFT? Or in other words: when can a particular HEFT be written in terms of the  $SU(2)$ doublet $\Phi$?  The answer is the 
following \cite{Alonso:2016oah}: given some four-dimensional HEFT scalar manifold with metric $g_{\alpha\beta}(\omega)$ 
(with $h=\omega^4$), it is possible to find a field reparametrization so that the Lagrangian can be written in terms 
of the doublet $\Phi$ whenever there exists a $SU(2)_L \times SU(2)_R$ invariant point on
the coset $G/H$. This is not always the case, but it happens in the SM. \myspace
The nonlinearity shows up as momentum-dependent vertices with an arbitrary high number of Goldstone boson insertions coming from
the expansion of the matrix field 
\begin{equation}\label{eq: Umatrix}
U=\exp\left(\frac{i\omega^a\sigma^a}{v}\right) \approx 1+i\frac{\omega^a\sigma^a}{v}+ O\left(\frac{\omega}{v}\right)^2
\end{equation} 
where $\omega=\{\omega^1,\omega^2,\omega^3\}$ and $\sigma^a$ represents the $SU(2)$ Pauli matrices. The range of validity of 
the HEFT
itself is given by the parameter controlling the expansion and sets a cutoff for the theory at $\Lambda = 4\pi v \approx 3$ TeV.
Given this, we would expect the resonances to emerge at a scale of a few TeV, values in principle reachable at the LHC, but
their detection is difficult by several reasons. The main one is that they are produced only in vector boson fusion, a process
that is subdominant at the LHC. The fact that the couplings of these putative resonances to the EWSBS are {\em a priori} unknown
is also a serious handicap. In addition, from previous work we know that the dynamical resonances in question are 
generically narrow
and not very visible, particularly if the anomalous couplings do not differ too much from their SM 
values \cite{Espriu:2012ih}. However, their
appearance is generic and by now their existence in extensions of the EWSBS seems well established by various
unitarization methods \cite{Delgado:2015kxa, Garcia-Garcia:2019oig}. Yet, detailed studies indicate that their confirmation 
may need the full 3000 fb$^{-1}$
statistics at the LHC~\cite{ Delgado:2017cls} if only leptonic decays ($4l$) are considered in the final states and about
one order of magnitude less if decays of the vector bosons in two jets are analyzed too ($2j+2l$) .\myspace
The terms of the chiral Lagrangian are organized by the chiral dimension of its local operators. This counts the number
of masses and derivatives (momenta) and a piece of the Lagrangian of chiral dimension $d$, $\mathcal{L}_d$, will contribute
to the process at order $\mathcal{O}(p^d)$. For our analysis with NLO precision, we restrict ourselves to operators up
to $\mathcal{O}(p^4)$. The set of operators that participate in (on-shell) $2\to 2$ scattering processes and are CP invariant, Lorentz
invariant, gauge and custodial symmetric are gathered in the following Lagrangians
\begin{equation}\label{eq: lag2}
\begin{split}
  \mathcal{L}_2 =&-\frac{1}{2g^2}\text{Tr}\left(\hat{W}_{\mu\nu}\hat{W}^{\mu\nu}\right)-
\frac{1}{2g^{\prime 2}}\text{Tr}\left(\hat{B}_{\mu\nu}\hat{B}^{\mu\nu}\right)
  +\frac{v^2}{4}\mathcal{F}(h)\text{Tr}\left(D^{\mu}U^{\dagger}D_{\mu}U\right)+\frac{1}{2}\partial_{\mu}h\partial^{\mu}h \\ 
  &-V(h) \vspace{0.2cm} \\
\end{split}
\end{equation}
\begin{equation}\label{eq: lag4}
\begin{split}
  \mathcal{L}_4 =&-i a_3\text{Tr}\left(\hat{W}_{\mu\nu}\left[V^{\mu},V^{\nu}\right]\right)
  +a_4 \left(\text{Tr}\left(V_{\mu}V_{\nu}\right)\right)^2
  +a_5 \left(\text{Tr}\left(V_{\mu}V^{\mu}\right)\right)^2+\frac{\gamma}{v^4}\left(\partial_{\mu}h\partial^{\mu}h\right)^2\\
  &+\frac{\delta}{v^2}\left(\partial_{\mu}h\partial^{\mu}h\right)\text{Tr}\left(D_{\mu}U^{\dagger}D^{\mu}U\right)
  +\frac{\eta}{v^2}\left(\partial_{\mu}h\partial_{\nu}h\right)\text{Tr}\left(D^{\mu}U^{\dagger}D^{\nu}U\right)\\
&+i\chi\,\text{Tr}\left(\hat{W}_{\mu\nu}V^{\mu}\right)\partial^{\nu}\mathcal{G}(h)
\end{split}
\end{equation}
with the usual definitions
\begin{equation}\label{eq: building_blocks}
\begin{split}
  &U=\exp\left(\frac{i\omega^a\sigma^a}{v}\right) \in SU(2)_V, \quad V_{\mu}=D_{\mu}U^{\dagger}U, 
  \quad \mathcal{F}(h)=1+2a\left(\frac{h}{v}\right)+b\left(\frac{h}{v}\right)^2+ \ldots , \\
  &D_{\mu}U=\partial_{\mu}U+i \hat{W}_{\mu} U, \quad \hat{W}_{\mu}=g\frac{\vec{W}_{\mu}\cdot\vec{\sigma}}{2},
  \quad \hat{W}_{\mu\nu}=\partial_{\mu}\hat{W}_{\nu}-\partial_{\nu}\hat{W}_{\mu}+i\left[\hat{W}_{\mu},\hat{W}_{\nu}\right], \\
  &V(h)=\frac{1}{2}M_h^2h^2+\lambda_3 v h^3+\frac{\lambda_4}{4}h^4+ \ldots,
  \quad \mathcal{G}(h)=1+ b_1\left(\frac{h}{v}\right)+b_2\left(\frac{h}{v}\right)^2+ \ldots
\end{split}
\end{equation}
From the last operator in (\ref{eq: lag4}), and taking into account the definitions above, we will just need for this study
the first term in the expansion of $\partial^{\nu}\mathcal{G}(h)$, so we define the new coupling $\zeta \equiv b_1 \chi$. 
In what concerns the Higgs potential $V(h)$, we will parametrize the departures from the SM trilinear and quartic self-couplings using the parameters $d_{3,4}$ such that $\lambda_{3,4}=d_{3,4}\lambda$, with $\lambda$ being the only SM Higgs 
self-interaction $\lambda=M_h^2/(2v^2)$ coupling. \myspace
The relevant HEFT for our processes up to NLO is then the sum of $\mathcal{L}_2$, $\mathcal{L}_4$ and the gauge fixing
and Faddeev-Popov terms
\begin{equation}\label{eq: lag_complete}
\mathcal{L}=\mathcal{L}_2+\mathcal{L}_4+\mathcal{L}_{GF}+\mathcal{L}_{FP}
\end{equation}
In the custodial limit and using an arbitrary gauge, the last two pieces are built using the following functions
\begin{equation}\label{eq: GFandFP}
\begin{split}
&f_i=\partial_{\mu}W_i^{\mu}-\frac{gv\xi}{2}\omega_i+\ldots\quad{i=1,2,3}\\
  &\mathcal{L}_{GF}= -\frac{1}{2\xi}\left(\sum_{i=1}^{3}f_i^2\right) \qquad\mathcal{L}_{FP}=
  \sum_{a,b=1}^{3}c_a^{\dagger}\frac{\delta f^{\prime}_a}{\delta \alpha_b}c_b
\end{split}
\end{equation}
where $f^{\prime}$ stands for the $SU(2)_L\times U(1)_{\mathcal{Y}}$ transformation of the function $f$ and $\alpha_a$
the gauge parameters.\myspace
We could enrich the HEFT with additional $SU(2)$ singlets $h^1(=h), h^2, h^3, \ldots$ and a term
\begin{equation}
  \frac12 g_{ij} \partial_\mu h^i \partial^\mu h^j - V(h^1, h^2,...).
\end{equation}
This possibility will not be considered here.
The interested reader can see \cite{Dobado:2019fxe} for more details.\myspace
The explicit gauge transformation is
\begin{equation}\label{eq: gauge_transforms}
   \hat{W}_{\mu}^{\prime}=g_L\hat{W}_{\mu}g_L^{\dagger}-\frac{1}{g}g_L\partial_{\mu}g_L^{\dagger} \qquad U^{\prime}=g_LU
\end{equation}
with  $g_L=\exp(i{\vec{\alpha}(x)\vec{\tau}}/{2})$,  $SU(2)$ matrix. The custodial transformation is
\begin{equation}
  U^\prime = g U g^\dagger,
\end{equation}
where $g$ is a constant $SU(2)$ matrix.\myspace
The notation we use roughly follows the conventions of \cite{Espriu:2014jya}. 
In \cite{Alonso:2012px} the reader may find a complete list of operators in the HEFT up to chiral dimension 
four~\footnote{Some operators are redundant~\cite{Buchalla:2013rka}, once  the bosonic  basis together with  
the fermionic one is considered and equations of motion are used.}. Only a subset
of those are relevant to us. Even taking this into consideration, the Lagrangians (\ref{eq: lag2}) and
(\ref{eq: lag4}) contain a number of free parameters. Not including these already well established by
experiments, we have: $a$ and $b$ in the $O(p^2)$ Lagrangian and $a_3, a_4, a_5, \gamma, \delta, \eta$
and $\zeta$ in the $O(p^4)$ Lagrangian. In the SM, $a=b=1$ and the rest are identically zero.
In addition, and of particular interest to us, we have $\lambda_3$ and $\lambda_4$ in the Higgs potential.
We expect departures from the SM values at most of order $10^{-3}$, possibly less,  in all the `anomalous' couplings. \myspace
The experimental situation concerning these couplings is as follows. The situation has been summarized e.g.
in \cite{Pich:2020xzo}. 
\begin{table}[tb]  
\begin{center}
\renewcommand{\arraystretch}{1.2}
\begin{tabular}{|c|c|c| }
\hline
Couplings & Ref. & Experiments\\ 
\hline \hline
$0.89<a<1.13$  & \cite{deBlas:2018tjm}
& LHC  \\ \hline 
$-0.76<b<2.56$ & \cite{ATLAS:2020jgy} & ATLAS \\ \hline
$-3.3\lambda<\lambda_3<8.5\lambda$ &\cite{CMS:2020tkr} & CMS \\ \hline
 $|a_1|<0.004$  & 
 \cite{Tanabashi:2018oca} & LEP ($S$-parameter)\\ \hline 
  $-0.06<a_2-a_3<0.20$&\cite{Almeida:2018cld} & LEP \& LHC  \\ \hline 
 $-0.0061<a_4 < 0.0063$&\cite{CMS:2019uys} & CMS (from $WZ\to 4l$) \\ \hline
$|a_5|<0.0008$ &  \cite{Sirunyan:2019der} & CMS (from $WZ/WW\to 2l2j$) \\  \hline
\end{tabular}
\caption{{\small
    Current experimental constraints on bosonic HEFT anomalous
    couplings at 95\% CL. See the text about the issue to extract the $a_4$ bound from the CMS 
analysis of \cite{Sirunyan:2019der}.}} \label{table: exp}
\end{center}
\end{table}
The experimental bounds for the chiral coupling $a$ have been measured by ATLAS and CMS in the subprocess
$h\to WW$ at 95\% C.L. to be $0.89<a<1.13$. Also, the first experimental bounds on the chiral parameter $b$
have been set by ATLAS with the subprocess $hh\to WW$. The result of this analysis, that assumes
the absence of new physics resonances, is $-1.02<b<2.71$. As we see, there are still large experimental
uncertainties regarding the Higgs couplings to vector bosons. These uncertainties affect operators of
chiral dimension two and are accordingly expected to be the most relevant ones. \myspace
The chiral couplings $a_4$ and $a_5$ have received a lot of attention in the past because to  a large
extent they control the appearance of resonances in the vector-isovector and scalar-isoscalar channels, at least
in the approximation were the ET is assumed to hold in its most strict version
and the propagation of transverse modes is neglected\footnote{Actually we will see below that the operators and 
couplings that survive in this extreme ET limit, i.e. $g=0$, {\em are} most relevant.}. Using
only the 8 TeV data, in 2017 ATLAS\cite{ATLAS:2016nmw} set the bounds $-0.024 < a_4 < 0.030$ and
$-0.028 < a_5 < 0.033$. More recently,
CMS~\cite{CMS:2019uys} using the 13 TeV data and only $4l$ decays from $WZ$ scattering was able to set the bounds
$-0.0061<a_4 < 0.0063$, $-0.0094 <  a_5 < 0.0098$, about three times better.
 In~\cite{Sirunyan:2019der}, CMS studies $2j+2l$ decays from both $WW$ and $WZ$ scatterings
to set the rather stringent   
bound\footnote{ CMS does not provide results for $a_4$ and $a_5$ directly as the analysis relies on the SMEFT, 
where the Higgs is treated as a doublet and the operators
contributing to the scattering of four $W$ are of dimension eight (unlike in the HEFT where they are of dimension four). 
The basis adopted is the one introduced in~\cite{Eboli:2006wa}, namely $f_{S,0}/\Lambda^4$ and $f_{S,1}/\Lambda^4$. 
However,  as was later noted in~\cite{Eboli:2016kko, Rauch:2016pai}, 
a third operator containing four derivatives of the Higgs doublet, with coefficient $f_{S,2}/\Lambda^4$, exists 
in the SMEFT and cannot be in general missed. In order
to get $f_{S,0}/\Lambda^4, f_{S,1}/\Lambda^4$ and $f_{S,2}/\Lambda^4$ one needs to measure $WW$ and  $WZ$  final states. 
This was done in the $4l$ analysis of CMS~\cite{CMS:2019uys}. However in~\cite{Sirunyan:2019der} $WW$ and $WZ$ are 
combined together and  it is not possible to extract
$f_{S,2}$ and $f_{S,0}$ separately. Note that only the sum of the operators corresponding to $f_{S,0}$ and $f_{S,2}$ 
is custodially invariant, but neither of them is. The sum matches the chiral
operator multiplying $a_4$~\cite{Rauch:2016pai} in the HEFT. Therefore, a valid
comparison requires assuming $f_{S,0}=f_{S,2}$ and only then
\bea
a_4= 
\frac{v^4}{8} \dfrac{f_{S,0}}{\Lambda^4}\Big\vert_{f_{S,2}=f_{S,0}}.\nonumber
\eea  
On the other hand, $f_{S,1}$ is custodially invariant and 
\bea
a_5= \frac{v^4}{16}\,\frac{f_{S,1}}{\Lambda^4}. \nonumber
\eea
}
$|a_5| < 0.0008$.  \myspace 
The smallness of these values in the case of $a_4$ and $a_5$ justifies, for the range of 
energies where they have been applied, the use of a simple
approach, without unitarization. Yet, small as they are, there is still room for new physics resulting from unitarization.
For instance this range still allows for the appearance of vector resonances in the range $ 1.5 $ TeV $< M_V < 2.5 $ TeV,
both for $a=1$ and $a=0.9$ \cite{ Delgado:2017cls}. 
\myspace
As for the coupling $a_3$, its range of uncertainty is quite large and its influence on the
location and properties of resonances in BSM physics has
not been assessed yet as this requires a full computation and subsequent unitarization studies including transverse
modes of the vector bosons. This will be presented below.\myspace
Concerning the Higgs potential parameters, there are not relevant bounds on $\lambda_4$ (i.e. on possible
departures from the SM relation $\lambda_4=\lambda = M_h^2/2 v^2$). 
In what concerns $\lambda_3$, and recalling the parametrization $\lambda_3=d_3\lambda$, some recent bounds have been 
obtained by ATLAS~\cite{ATLAS:2019pbo}, $-2.3<d_3<10.3$, combining double and single- Higgs analysis at 95\% C.L.,  
and by CMS~\cite{CMS:2020tkr}, $-3.3<d_3<8.5$, from the subprocess $HH\to b\bar{b}\gamma\gamma$ .
To our knowledge, there are  no experimental studies on
the $\mathcal{O}(p^4)$ chiral parameter $\zeta$. However, as we stress in our work, this parameter plays a 
role in the $WW$-scattering at one loop.

\section{Tree level calculation of the relevant $2 \to 2$ processes}\label{sec:tree-level}

As mentioned in the introduction in order to implement a fair comparison with experiment,
we are interested in obtaining unitary amplitudes for the following $2\to 2 $
processes with one loop precision: $W_L W_L \to W_L W_L $, $ W_L W_L \to hh$ and $hh \to hh$. In the first case,
the $I=0, J=0$ (weak) isospin and angular momentum projection will be of most interest to us, but we will
actually provide results that can be used for any $I,J$ projection thanks to the relations resulting from the
exact isospin symmetry present for $g^\prime=0$. For instance, provided that custodial symmetry remains exact, 
from the $\wplus \wminus \to ZZ$ amplitude it is possible to obtain all the remaining
$W^a_LW^b_L \to W^c_LW^d_L$ ones thanks to the isospin relations (see e.g. \cite{Espriu:2012ih} for details).
From Bose and crossing symmetries,
\begin{equation}\label{eq: isos_general}
\mathcal{A}^{abcd}=\delta^{ab}\delta^{cd}\mathcal{A}\left(p^a,p^b,p^c,p^d\right)
+\delta^{ac}\delta^{bd}\mathcal{A}\left(p^a,-p^c,-p^b,p^d\right)+\delta^{ad}\delta^{bc}\mathcal{A}\left(p^a,-p^d,p^c,-p^b\right)
\end{equation}
which allows us to write
\begin{equation}\label{eq: isos_I1}
\begin{split}
\mathcal{A}^{+-00}&=\mathcal{A}(p^a,p^b,p^c,p^d)\\
\mathcal{A}^{+-+-}&=\mathcal{A}(p^a,p^b,p^c,p^d)+\mathcal{A}(p^a,-p^c,-p^b,p^d)\\
\mathcal{A}^{++++}&=\mathcal{A}(p^a,-p^c,-p^b,p^d)+\mathcal{A}(p^a,-p^d,p^c,-p^b)
\end{split}
\end{equation}
This means that every amplitude with vector bosons as asymptotic states can be obtained by crossings 
from the fundamental amplitude $W^{+}W^{-}\to ZZ$, as mentioned before. Notice that crossing when longitudinally-polarized gauge bosons are involved has to be implemented via the momenta and not via Mandelstam variables
because the polarization vectors do not transform covariantly (see e.g., the discussion in \cite{Espriu:2012ih}). \myspace
The fixed isospin projections $T_I$ are given by
\begin{equation}\label{eq: Ts}
\begin{split}
T_0&=3\mathcal{A}^{+-00}+\mathcal{A}^{++++}\\
T_1&=2\mathcal{A}^{+-+-}-2\mathcal{A}^{+-00}-\mathcal{A}^{++++}\\
T_2&=\mathcal{A}^{++++}
\end{split}
\end{equation}
Taking into account that in this framework the Higgs is a singlet, we can also write the projections 
for the crossed channels with an $I=0$ external state and the corresponding isospin amplitudes
\begin{equation}\label{eq: isos_I0}
\mathcal{A}(W_L^a W_L^b \to hh)=\mathcal{A}^{ab}(p^a,p^b,p_{h,1},p_{h,2}),\qquad
T_{Wh,0}=\sqrt{3}\mathcal{A}^{+-},
\end{equation}
\begin{equation}
\mathcal{A}(hh\to hh)=\mathcal{A}(p_{h,1},p_{h,2},p_{h,3},p_{h,4})= T_{hh,0},
\end{equation}
where the last amplitude has obviously only an $I=0$ projection.\myspace
All the tree-level amplitudes gathered above have both LO (computed using the Feynman rules from $\mathcal{L}_2$) 
and the NLO contributions (obtained using the rules of $\mathcal{L}_4$).\myspace
Below we present the tree level amplitudes for the different $2\to2$ processes that are important 
for our study. We use the following notation: a superindex indicates the different processes labeled 
as $WW$ for $W^+W^-\to ZZ$, $Wh$ for $W^+W^-\to hh$ and $hh$ for $hh\to hh$. Also, each amplitude carries a 
subindex ${xy}$ that represents a process with a particle $y$ propagating in the $x$ channel. 
In the case with $x=c$ and no $y$, $\mathcal{A}_{c}$, represents the contact interaction of the four external particles. 
For instance, the amplitude $\mathcal{A}^{WW}_{sh}$ represents a Higgs exchanged in the $s$-channel of $W^+W^-\to ZZ$ scattering.

\subsection{$\wplus \wminus \to ZZ$}
The tree-level amplitude includes contribution from the $\mathcal{O}(p^2)$ and $\mathcal{O}(p^4)$ Lagrangian
\begin{equation}\label{eq: tree_WW}
\begin{split}
\mathcal{A}^{WW}_c=&g^2\left(\left((-2 a_3+a_4)g^2+1\right)\left(\left(\varepsilon_1\varepsilon_4\right) \left(\varepsilon_2\varepsilon_3\right)+ \left(\varepsilon_1 \varepsilon_3\right) \left(\varepsilon_2 \varepsilon_4\right)\right)\right.\\
&\left.+2 \left((2 a_3+a_5)g^2-1\right) (\varepsilon_1\varepsilon_2) \left(\varepsilon_3\varepsilon_4\right)\right)\\
\mathcal{A}^{WW}_{sh}=&-\frac{a^2g^2M_W^2\left(\varepsilon_1\varepsilon_2\right)\left(\varepsilon_3\varepsilon_4\right)}{(p_1+p_2)^2-M_H^2}+\frac{ag^4\zeta}{4((p_1+p_2)^2-M_H^2)}\left[2(\varepsilon_3\varepsilon_4)\left((p_1\varepsilon_2)(p_2\varepsilon_1) \right.\right.\\
&\left.\left. -(\varepsilon_1\varepsilon_2)(p_1+p_2)^2\right)+2(\varepsilon_1\varepsilon_2)(p_3\varepsilon_4)(p_4\varepsilon_3)\right]\\
\mathcal{A}^{WW}_{tW}=&-\frac{(1-2a_3g^2)g^2}{(p_1-p_3)^2-M_W^2} \left[-4\left((\varepsilon_1\varepsilon_2)(p_1\varepsilon_3)(p_2\varepsilon_4)+(\varepsilon_1\varepsilon_4)(p_1\varepsilon_3)(p_4\varepsilon_2)\right.\right.\\
&\left. +(\varepsilon_2\varepsilon_3)(p_3\varepsilon_1)(p_2\varepsilon_4)+(\varepsilon_3\varepsilon_4)(p_3\varepsilon_1)(p_4\varepsilon_2) \right)\\&+2\left( (\varepsilon_2\varepsilon_4)\left( (p_1\varepsilon_3)(p_2+p_4)\varepsilon_1+(p_3\varepsilon_1)(p_2+p_4)\varepsilon_3\right)\right.\\
&\left. +(\varepsilon_1\varepsilon_3)((p_2\varepsilon_4)(p_1+p_3)\varepsilon_2+(p_4\varepsilon_2)(p_1+p_3)\varepsilon_4)\right)\\
&\left.-(\varepsilon_1\varepsilon_3)(\varepsilon_2\varepsilon_4)((p_1+p_3)p_2+(p_2+p_4)p_1) \right]\\
\mathcal{A}^{WW}_{uW}=&\mathcal{A}_{tW}(p_3\leftrightarrow p_4, \varepsilon_3 \leftrightarrow \varepsilon_4)
\end{split}
\end{equation}
where $\varepsilon_i$ is the abbreviation for $\varepsilon_L(p_i)$.

\subsection{$W_L W_L \to hh$}
\begin{equation}\label{eq: tree_Wh}
\begin{split}
\mathcal{A}^{Wh}_c=&\frac{g^2\,b}{2}(\varepsilon_1\varepsilon_2)-\frac{g^2\,\eta}{v^2}((\varepsilon_1 p_4)(\varepsilon_2 p_3)+(p_3\varepsilon_1)(\varepsilon_2 p_4))-\frac{2g^2\,\delta}{v^2}(p_3 p_4)(\varepsilon_1\varepsilon_2)\\
&+\frac{g^2\,\zeta}{v^2}((\varepsilon_1\varepsilon_2)(p_1+p_2)^2-2(p_1\varepsilon_2)(p_2\varepsilon_1))\\
\mathcal{A}^{Wh}_{sh}=&\frac{3g^2M_h^2}{2((p_1+p_2)^2-M_h^2)}\left(a(\varepsilon_1\varepsilon_2)+\frac{\zeta}{v^2}((\varepsilon_1\varepsilon_2)(p_1+p_2)^2-2(p_1\varepsilon_2)(p_2\varepsilon_1))\right)\\
\mathcal{A}^{Wh}_{t\omega}=&\frac{2a^2g^2+a\zeta g^4}{2(p_1-p_3)^2}\left((p_3\varepsilon_1)(p_4\varepsilon_2)\right)\\
\mathcal{A}_{tW}=&\frac{a^2g^2M_W^2}{((p_1-p_3)^2-M_W^2)}\left(\varepsilon_1\varepsilon_2+\frac{(p_4\varepsilon_2)(\varepsilon_1 p_3)}{(p_1-p_3)^2}\right)+\frac{ag^4\zeta}{2((p_1-p_3)^2-M_W^2)}\left( 2M_h^2(\varepsilon_1\varepsilon_2)\right. \\
&\left. -(p_4\varepsilon_2)(p_2\varepsilon_1)-(\varepsilon_1p_3)(\varepsilon_2p_3)+M_W^2\frac{(p_4\varepsilon_2)(\varepsilon_1p_3)}{(p_1-p_3)^2}\right)\\
\mathcal{A}^{Wh}_{u\omega}=&\mathcal{A}^{Wh}_{t\omega}(p_3\leftrightarrow p_4)\\
\mathcal{A}^{Wh}_{uW}=&\mathcal{A}^{Wh}_{tW}(p_3\leftrightarrow p_4)\\
\end{split}
\end{equation}

\subsection{$hh \to hh$}
\begin{equation}\label{eq: tree_hh}
\begin{split}
\mathcal{A}^{hh}_c=&\frac{8\gamma}{v^4}\left((p_1 p_4)(p_2 p_3)+(p_1 p_3)(p_2 p_4)+(p_1 p_2)(p_3 p_4)\right)-6\lambda_4\\
\mathcal{A}^{hh}_{sh}=&-\frac{36\lambda_3^2v^2}{(p_1+p_2)^2-M_h^2}\\
\mathcal{A}^{hh}_{th}=&\mathcal{A}^{hh}_{sh}(p_2\leftrightarrow -p_3)\\
\mathcal{A}^{hh}_{uh}=&\mathcal{A}^{hh}_{sh}(p_2\leftrightarrow -p_4)
\end{split}
\end{equation}

\subsection{Counterterms}
The divergences eventually appearing in all these processes at the one-loop level have to be absorbed by redefining
the parameters appearing at tree level. Namely,
\begin{equation}\label{eq: renor_couplings}
\begin{split}
&v^2\rightarrow v^2+\delta v_{div}^2+\delta \bar{v^2}, \quad \{h,\omega\} \rightarrow Z_{h,\omega}\{h,W,\omega\}, 
\quad M^2_{h,W}\rightarrow M^2_{h,W}+\delta M^2_{h,W}, \\ 
&\lambda_{3,4}\rightarrow \lambda_{3,4}+\delta \lambda_{3,4}, \quad a\rightarrow a+\delta a, \quad b\rightarrow b+\delta b, \quad a_i \rightarrow a_i+\delta a_i,\\ 
&\delta \rightarrow \delta + \delta \delta, \quad \eta \rightarrow \eta + \delta \eta, \quad \gamma \rightarrow \gamma 
+ \delta \gamma, \quad \zeta \rightarrow \zeta + \delta \zeta
\end{split}
\end{equation}
where we recall that $\zeta \equiv\chi b_1$.\myspace
Even though the gauge coupling $g$ appears in some of the previous formulae, the relation
$M_W= gv/2$ is assumed to all orders and the renormalization of $g$ is fixed by the ones of $v$ and $M_W$. 
On the contrary, we cannot assume the SM relation $M_h^2 = 2 v^2 \lambda$ because this already assumes
the persistence of the SM Higgs potential -something that we want to eventually test. It is for this reason that we keep separate notations $\lambda_3$ and $\lambda_4$ for the three- and four-point Higgs vertices.\myspace
In general all counterterms have both a divergent and a finite part, determined by the renormalization conditions. However,
for reasons that will be clear later, we have split the counterterms for $v^2$ explicitly into divergent and
finite pieces.\myspace
As we will see subsequently, we will determine all counterterms for processes involving only Goldstone bosons, whose
calculation is substantially simpler than using vector bosons. This is enough to get all the necessary counterterms.
The corresponding tree-level amplitudes for the
Goldstones will be given in the next section.\myspace

\subsection{Auxiliary processes: $h\to\omega\omega$, $h\to hh$ and $h\to W\omega$}
In this subsection we collect a series of $1\to 2$ tree-level processes that are useful to uniquely determine 
the counterterms. One of the processes ($h \to hh$) cannot take place on shell, but it has to be rendered finite through the renormalization procedure. They  are
\begin{enumerate}
\item \textbf{$h\to\omega\omega$ process}.\\
The tree-level amplitude of this decay up to NLO is (with $p_h$  the Higgs 4-moment) 
\begin{equation}\label{eq: decay_hww}
\mathcal{A}^{h\to \omega\omega}_{tree}=-\frac{a p_h^2}{v}
\end{equation}
which leads to the on-shell renormalization condition
\begin{equation}\label{eq: count_hww}
\frac{M_h^2}{2v^3}\left( a\delta v^2-2v^2\delta a\right)+\text{div}\left(\mathcal{A}^{h\to\omega\omega}_{1-loop}\right)=0.
\end{equation}
From (\ref{eq: decay_hww}) and with the substitutions that will be specified later, we find the relation
between $\delta a$ and $\delta v^2$,
being the counterterms associated to the chiral parameter $a$ and to the vev, respectively. Note
that obviously $p_h$ does no get a counterterm even though on shell $p_h^2 = M_h^2$. 
\item\textbf{$h\to hh$ process}.\\
At tree level, the corresponding amplitude reads
\begin{equation}\label{eq: decay_hhh}
\mathcal{A}^{h\to hh}_{tree}=-6\lambda_3v\,.
\end{equation}
From the cancellation of the divergences of this process at one loop, we get  a relation between $\delta \lambda_3$ 
and $\delta v^2$
\begin{equation}\label{eq: count_hhh}
-\frac{3}{2v^3}\left(d_3 M_h^2\delta v^2+4v^4\delta \lambda_3 \right)+\text{div}\left(\mathcal{A}^{h\to hh}_{1-loop}\right)=0.
\end{equation}
Note that this (off-shell) process cannot be modified by using the equation of motion for $h$.

\item \textbf{$h\to W\omega$ process}
\begin{equation}
\mathcal{A}^{h\to W\omega}_{tree}= ig\left(a+\frac{M_W^2\zeta}{v^2}\right) \varepsilon_W p_h\,.
\end{equation}
From the cancellation of the divergences of this process at one loop level and with the assumption that
the relation $M_W=\frac{1}{2}gv$
is satisfied at every order, we obtain a relation among $\delta v^2,\, \delta M_W^2,\, \delta a$ and $\delta\zeta$.
\begin{equation}\label{eq: count_hWw}
  -i\left(a M_W^2\delta v^2-2M_W^4\delta\zeta-av^2\delta M_W^2-2 M_W^2 v^2 \delta a\right)\frac{\varepsilon_W p_h}{M_W v^3} 
  +\text{div}\left(\mathcal{A}^{h\to W\omega}_{1-loop}\right)=0.
\end{equation}
\end{enumerate}

\section{One loop calculation of the relevant $2 \to 2$ processes and counterterms}\label{sec:loop-level} 
In this section we present the one loop calculation of the relevant amplitudes. 
The amplitudes cannot be expressed in terms of elementary functions as they are given by Passarino-Veltman
integrals and they are quite cumbersome. For this reason we just show here the divergent parts (only present in the
real part of the amplitude) and the explicit expression for the counterterms.\myspace
The calculation of quantum corrections for the processes requires gauge fixing and the inclusion of the Faddeev-Popov ghosts.
The results presented below will be given in the Landau gauge $\xi=0$. Obviously, physical amplitudes should be independent
of the gauge choice, but some renormalization constants do depend on the gauge election. In the gauge-fixing processes
several differences are present in the HEFT~\cite{Herrero:1993nc, Herrero:2020dtv} with respect to the textbook SM. 
On one hand, the Higgs is a singlet so it does not
play any role in the symmetry and thus it is not present either in the gauge fixing piece, or in the Faddeev-Popov one,
i.e, there are no Higgs-ghost interactions. On the other hand, the gauge condition in (\ref{eq: GFandFP}) translates
into ghost-antighost pairs coupled to an arbitrary number of Goldstones insertions with a strength depending
on the gauge parameter.\myspace
As mentioned before, the calculation of the  $2\to 2$ amplitudes we are interested in is relatively involved, 
even at the one-loop level. Recall that we will be interested both in the divergent part (to determine counterterms) 
but also in the much more involved finite part. This is particularly so because there are several free parameters 
that have to be considered when one moves away from the SM. For this reason it has become customary starting with 
the work of \cite{Espriu:2012ih} to split the one-loop calculation into two parts. The imaginary part is computed
exactly from the tree-level results described in Sec.\ref{sec:tree-level} using the optical theorem, including only the
$\mathcal{O}(p^2)$ pieces. The real part is computed making use of the equivalence theorem, replacing
the longitudinal vector
bosons in the external legs with the corresponding Goldstone bosons. However, the full set of polarizations (including of
course transverse modes) will be kept internally inside the loops in the present study. We emphasize that this procedure is
done only for efficiency reasons and there is no fundamental reason to do so.\myspace
Two reasons for concern might arise if this splitting between the real and imaginary parts is used. The first one is whether
this actually preserves unitarity for unitarized amplitudes. A reassuring check will be presented in Sec.\ref{sec:unitarization},
but the verification is actually guaranteed because the ET is working quite accurately provided that $s \gg M_W^2$, which is
the regime we are actually interested. In any case, unitarity should not be confused with the concept of
'perturbative unitarity' that relates the real and imaginary part up to a given order in perturbation theory and that only
implies the consistency of the calculation in a field theory (even if this theory is nonunitary) and it is therefore
automatic and of no interest to us; we have nevertheless verified that perturbative unitarity is well reproduced at the
level of a few per cent as a check of the calculation by comparing the imaginary parts obtained in either way.\myspace 
A second concern could be whether gauge invariance 
is preserved by doing this splitting. The answer is obviously in the affirmative in the following sense.
The ET is derived from gauge invariance by requiring that in and out states fulfill the gauge condition
(see e.g. \cite{Espriu:1994ep}). A precise implementation of the ET tells us that corrections to the
leading term (i.e. the one where the longitudinal gauge boson amplitude is approximated by the corresponding Goldstone boson
scattering) are given by a succession of subleading contributions each one lower with respect to the previous
by a power of momenta\footnote{The ET relies on the splitting of the polarization vector $\epsilon^\mu_L = k^\mu/M_W + v^\mu$. Here
  $v^\mu$ is of order $M_W/E$. Substituting the splitting into the amplitude leads to corrections with  higher and higher
  powers of $E$ in the denominator. When summed up they all reproduce the original $W_L$ amplitude. The reader can see
  \cite{Espriu:1994ep} for details. We note that the ET is used here for the one-loop correction only, not for the tree
  level contribution - different orders of $\hbar$. The one-loop correction to the partial wave is of $O(s^2)$ and
  the corrections implied by the ET might change the $O(s)$ contribution, but the latter - tree level- is calculated exactly
  without appealing to the ET. Therefore gauge invariance is respected even if the splitting is itself not gauge invariant.}.
 When translated into partial waves this implies that each of these successive corrections
is suppressed by one more power of $s$. Taking into account that the one-loop amplitude is nominally of order $s^2$,
corrections might change the $O(s)$ contribution only. However, the $O(s)$ contribution to the amplitude
is computed exactly, without appealing to the ET. Therefore
gauge invariance is guaranteed at the order we are computing.
Having said that, it is safe to use the 't Hooft-Landau gauge which considerably simplifies the calculation. Where a comparison can be made, all counterterms agree with those computed in a general gauge 
as we will see below.\myspace

\subsection{Real part: The equivalence theorem}
The ET states that at high energies compared to the electroweak scale, the longitudinal projection of
the vector boson can be substituted by the associated Goldstone boson allowing an error 
\begin{equation}\label{eq: et_pol}
\varepsilon^{\mu}_L(k)=\frac{k^{\mu}}{M_W}+\mathcal{O}\left(\frac{M_W}{\sqrt{s}}\right).
\end{equation}
This error assumed at the TeV scale, the cutoff of our theory, is then, nominally, lower than 10\% but actually much lower because $M_W$ can appear only quadratically.\myspace
The calculation carried out in Ref. \citep{Espriu:2013fia} just allowed the longitudinal part of the gauge bosons running inside
the loops but for this study a full $\mathcal{O}(g)$ calculation is performed and the number of diagrams that needs to be
taken into account scales to more that 1500. This calculation has been done with the help of FeynArts~\cite{Hahn:2000kx},
FeynCalc~\cite{Shtabovenko:2020gxv} and FeynHelpers~\cite{Shtabovenko:2016whf} Mathematica packages.
These routines are able to evaluate the one-loop integrals in the Passarino-Veltman notation\cite{Passarino:1978jh} 
and extract just the divergent part of the diagrams when is required.\myspace
The expressions (\ref{eq: isos_general})-(\ref{eq: isos_I0}) are also valid within the equivalence theorem but now the symmetry
will be manifest at the level of the Mandelstam variables themselves in the absence of polarization vectors that do not transform
as four vectors under Lorentz transformations \citep{Espriu:2014jya}, which is a nice simplification.\myspace
After use of the ET we have to consider the (real part of) the following processes.

\subsubsection{$\omega^+\omega^-\to zz$}
From the isospin point of view, this is the fundamental amplitude for elastic $\omega\omega$ scattering. In this process
294 1PI diagrams participate at one-loop level. The divergences that appear need to be absorbed by redefinitions of
coefficients of the tree-level amplitude up to NLO. When the $W_L$ are replaced by the $\omega$, following
the equivalence theorem, the amplitude tree-level amplitude reads
\begin{equation}\label{eq: tree_ww}
\begin{split}
  \mathcal{A}^{\omega\omega}_{tree}=& -\frac{s(M_h^2-s(1-a^2))}{(s-M_h^2)v^2}+\frac{4}{v^4}(a_4(t^2+u^2)+2a_5s^2)+
  \left[\frac{g^2}{4}\frac{u-s}{t-M_W^2}\left(1+\frac{8a_3t}{v^2}\right) \right.\\
  &\left.+ u\Longleftrightarrow t \right]
\end{split}
\end{equation} 
with the infinitesimal substitutions
\begin{equation}\label{eq: counters_wwww}
\begin{split}
&M_h^2\to M_h^2+\delta M_h^2, \qquad M_W^2\to M_W^2+\delta M_W^2, \qquad v^2\to v^2+\delta v^2,\\
&a\to a+\delta a, \qquad a_4\to a_4+\delta a_4, \qquad a_5\to a_5+\delta a_5, \qquad a_3\to a_3+\delta a_3
\end{split}
\end{equation}
Besides, a redefinition of the Goldstone fields in the Lagrangian needs to be used for the divergent
corrections of the external legs
\begin{equation}\label{eq: Z_w}
\{\omega^{\pm},z\}\to \sqrt{Z_{\omega^{\pm},z}}\{\omega^{\pm},z\} \approx (1+\frac{1}{2}\delta Z_{\omega^{\pm},z} )\{\omega^{\pm},z\}
\end{equation}

\subsubsection{$\omega^+\omega^-\to hh$}
This scattering requires computing 505 one-loop 1PI diagrams . The tree-level amplitude is
\begin{equation}\label{eq: tree_wh}
\begin{split}
  \mathcal{A}^{\omega h}_{tree}=&-b\frac{s}{v^2}-\frac{6a\lambda_3 s}{s-M_h^2}
  -\left[\frac{g^2}{4(t-M_W^2)}\left(2a^2s+\frac{a^2}{t}(t-M_h^2)^2\right)+t \Longleftrightarrow u\right]\\
&-\frac{1}{v^2}\left[\frac{\zeta ag^2}{2(t-M_W^2)}\left(t(s-u)+
M_h^4\right)+ t\Longleftrightarrow u\right]-\frac{1}{v^2}\left[\frac{a^2}{t}(t-M_h^2)^2\right.\\
&\left.+t\Longleftrightarrow u\right]+\frac{1}{v^4}\left(2\delta\, s(s-M_h^2)+\eta\left((t-M_h^2)^2+(u-M_h^2)\right)\right)
\end{split}
\end{equation}
To get rid of the divergences of this process, the following substitutions for the couplings are needed
\begin{equation}\label{counters_ wwhh}
\begin{split}
&M_h^2\to M_h^2+\delta M_h^2, \qquad M_W^2\to M_W^2+\delta M_W^2, \qquad v^2\to v^2+\delta v^2, \qquad a\to a+\delta a\\
  &b\to b+\delta b, \qquad \lambda_3\to \lambda_3+\delta \lambda_3, \qquad \delta\to \delta+\delta\, \delta,
  \qquad \eta\to \eta+\delta\eta, \qquad \zeta\to\zeta+\delta\zeta 
\end{split}
\end{equation}
Now, apart from (\ref{eq: Z_w}), we will also need the redefinition of the classical Higgs field 
\begin{equation}\label{eq: Z_h}
h\to \sqrt{Z_h}h \approx (1+\frac{1}{2}\delta Z_h)h
\end{equation}
\subsubsection{$hh\to hh$}
This process at the one-loop level contains 654 1PI diagrams and the divergences must be canceled from the
parameters of the amplitude (\ref{eq: tree_hh}) once the usual Mandelstam definitions have been applied,\myspace
\begin{equation}\label{eq: tree_hh_2}
\begin{split}
\mathcal{A}_{tree}^{hh}=&-6\lambda_4-36\lambda_3^2v^2\left(\frac{1}{s-M_h^2}+\frac{1}{t-M_h^2}+\frac{1}{u-M_h^2}\right)+\frac{8\gamma}{v^4}\left(\left(\frac{s}{2}-M_h^2\right)^2\right.\\
&\left.+\left(\frac{t}{2}-M_h^2\right)^2+\left(\frac{u}{2}-M_h^2\right)^2 \right)
\end{split}
\end{equation}
The universal counterterms
\begin{equation}
\begin{split}
  &M_h^2\to M_h^2+\delta M_h^2, \quad v^2\to v^2+\delta v^2, \quad \lambda_3\to \lambda_3+\delta \lambda_3,\\
  &\qquad \quad \lambda_4\to \lambda_4+\delta \lambda_4, \quad \gamma\to \gamma+\delta \gamma
\end{split}
\end{equation}
are required for absorbing the divergences, plus the Higgs redefinition (\ref{eq: Z_h}).

\subsection{Determination of counterterms}

The real absorptive part has both finite and divergent parts. The divergences are reabsorbed
in the amplitudes via new parameters from redefinitions of couplings and fields of the bare
theory (\ref{eq: lag_complete})  given in the previous subsections.\myspace
The counterterms of our theory are not uniquely defined and depend on the choice
of physical inputs to define the finite part of the amplitude. In this study the so-called on-shell 
(OS) scheme (see e.g. \cite{Grozin:2005yg})
has been used. It states that the physical mass is placed in the pole of the renormalized propagator with residue 1.
This means 
\begin{equation}\label{eq: renor_prop}
\begin{split}
  Re\left[\Pi_{h,W_T}(q^2=M_{h,W_T}^2)-\delta M_{h,W_T}^2\right]=0, \quad Re\left[\frac{d\Pi_{h,\omega,W}}{dq^2}(q^2=M^2_{h,W,\omega})
    +\delta Z_{h,W,\omega}\right]=0
\end{split}
\end{equation}
where $\Pi(q^2)$ is the one-loop correction to the respective propagator. The OS, first used in the context of LEP
physics, has the advantage that many relevant
radiative corrections involve only two-point functions. This is obvious for the masses and wave function renormalization. 
After the splittings $\delta M_{h,W}^2 = \delta \overline{M}_{h,W}^2+\delta M_{h,W,div}^2$ and $\delta Z_{h,\omega}=\delta \bar{Z}_{h,\omega}+\delta Z_{h,\omega,div}$ we obtain 
\begin{equation}\label{eq: counter_masses_Zs}
\begin{split}
&\delta M_{h,div}^2=\frac{\Delta}{32 \pi^2 v^2}\left(3\left[6\left(2a^2+b\right)M_W^4-6a^2M_W^2M_h^2+\left(3d_3^2+d_4+a^2\right)M_h^4 \right]\right),\\
&\delta M_{W,div}^2=\frac{\Delta}{48 \pi^2 v^2}\left(M_W^2\left[3\left(b-a^2\right)M_h^2+\left(-69+10a^2\right)M_W^2\right]\right),\\
&\delta Z_{h,div}=\frac{\Delta}{16 \pi^2 v^2}\left(3 a^2\left(3 M_W^2-M_h^2 \right)\right),\\
&\delta Z_{\omega,div}=\frac{\Delta}{16 \pi^2 v^2}\left(\left(b-a^2\right)M_h^2+3 \left(a^2+2\right) M_W^2\right)
\end{split}
\end{equation}
where $\Delta \equiv \frac{1}{\epsilon}+log(4\pi)+\gamma_E$ and the dimensionality is set to $4+2\epsilon$.\myspace
The one-loop level propagator mixing between the gauge boson and its associated Goldstone is protected by the gauge
fixing condition in (\ref{eq: GFandFP}) and no extra counterterms will be needed for this. In the absence
of electromagnetic interactions assuming an exact custodial symmetry, no $Z-\gamma$ mixing in the gauge
propagator can occur either.\myspace
Besides, the condition of vanishing tadpole is assumed. There is an extra counterterm $\delta T$ that cancels the Higgs tadpole contribution at one loop satisfying the usual relation \cite{Espriu:2013fia}
\begin{equation}\label{eq: tadpole_counter}
\delta T= -v\left(\delta M_h^2-2v^2\delta\lambda-2\lambda\delta v^2\right)=-\mathcal{A}_{tad}^h
\end{equation} 
With our parametrization for the Higgs potential, $\lambda$ does not appear in any of the processes but its counterterm can be determined using (\ref{eq: tadpole_counter}) once $\delta M_h^2$ and $\delta v^2$ are obtained. \myspace
The matrix field (\ref{eq: Umatrix}) containing the Goldstones in the HEFT should retain its unitarity and hence
it cannot receive any multiplicative renormalization. Perturbatively, the redefinitions of the $n-th$ term
of the expansion of $U$
\begin{equation}\label{eq: expansion_U_term}
  \frac{1}{n!}\left(i\frac{\omega}{v}\right)^n \to \frac{1}{n!}\left(i\frac{\omega}{v}\right)^n
  +\frac{1}{2(n-1)!}\left(\delta Z_{\omega}-\frac{\delta v^2}{v^2}\right)\left(i\frac{\omega}{v}\right)^n
\end{equation}
It turns out that to absorb the one-loop divergences, the counterterms for the Goldstone
fields ($\sqrt{Z_\omega}$) and the vev ($\sqrt{\delta v^2}$) are equal
so they cancel each other at every order in the expansion. The
{\em finite} part of $\sqrt{\delta v^2}$ is fixed, at every order, by the condition
\begin{equation}\label{eq: relation_Z_v}
\delta Z_{\omega}=\frac{\delta v^2}{v^2}\,.
\end{equation}
The counterterms of the HEFT whose renormalization is not determined by the OS scheme conditions, are obtained in the $\overline{MS}$ scheme. Since this is a mass
independent scheme, the counterterms corresponding to operators of dimension four (such as $a_3$, $a_4$, etc. ) are independent of $M_W$. \myspace
The complete list of counterterms allowing us to get rid of the divergences of the three amplitudes in the
previous subsections is 
\begin{equation}\label{eq: counter_coupling}
\begin{split}
&\delta v^2_{div}=\frac{\Delta}{16\pi^2}\left((b-a^2)M_h^2+3(a^2+2)M_W^2\right),  \quad \delta T_{div}=-\frac{\Delta}{32\pi^2 v}3\left(d_3M_h^4+6aM_W^4\right),\\
  &\delta a=\frac{\Delta}{32 \pi ^2 v^2}\left(6\,a \left(-2 a^2+b+1\right)M_W^2+(5a^3-a(2+3b)-3d_3(a^2-b))M_h^2\right), \\
&\delta b=\frac{\Delta}{32 \pi ^2 v^2}\left(6  \left(3 a^4-6 a^2 b+b (b+2)\right)M_W^2 \right. \\
&\qquad\left. -\left(21a^4-a^2(8+19b)+b(4+2b)+6ad_3(1+2b-3a^2)-3d_4(b-a^2)\right)M_h^2\right), \\
&\delta \lambda_{div}= \frac{\Delta}{64 \pi ^2 v^4}\left(\left(5a^2-2 b+3\left(d_3(3d_3-1)+d_4\right)\right)M_h^4 -12 \left(2 a^2+1\right) M_W^2 M_h^2\right.\\
&\qquad\left.+18  (a (2a-1)+b) M_W^4\right), \\
&\delta \lambda_3= \frac{\Delta}{64 \pi ^2 v^4}\left(36a b M_W^4+6 (3a^3-3ab-d_3(5a^2+1))M_W^2 M_h^2  \right. \\
&\qquad\left. +(-9a^3+3ab+d_3(10a^2-b)+9d_3d_4)M_h^4 \right), \\
&\delta \lambda_4=\frac{\Delta}{64\pi^2v^4}\left(36b^2M_W^4-12(a^2-b)(8a^2-2b-9ad_3)M_W^2M_h^2 \right.\\
&\qquad\left.+(96a^4+4b^2-d_3(114a^3-42ab)+9d_4^2+a^2(-64b+27d_3^2+12d_4))M_h^4 \right) ,\\
  &\delta a_3=-\frac{\Delta}{384\pi^2}\left(1-a^2\right),\quad \delta a_4=-\frac{\Delta}{192 \pi ^2}\left(1-a^2\right)^2, \\
 &\delta a_5=-\frac{\Delta}{768 \pi ^2}\left(5 a^4-2 a^2 (3b+2)+3 b^2+2\right),\\
  &\delta \gamma=-\frac{\Delta}{64\pi^2}3(b-a^2)^2, \quad \delta \delta = -\frac{\Delta}{192\pi^2} (b-a^2)(7a^2-b-6),
  \quad \delta \eta=-\frac{\Delta}{48\pi^2} (b-a^2)^2, \\
&\delta \zeta=\frac{\Delta}{96\pi^2}a(b-a^2)\,.
\end{split}
\end{equation}
For completeness we include the counterterm
for $\delta g^2$ even though this is not an independent input of the theory anymore in the renormalization scheme used here.
\begin{equation}\label{eq: counter_g}
\delta g^2=g^2\left(\frac{\delta M_W^2}{M_W^2}+\frac{\delta v^2}{v^2}\right)=\frac{\Delta}{12\pi^2v^4}M_W^2\left((-51+19a^2)M_W^2+6(b-a^2)M_h^2\right)
\end{equation}
Notice that our prescription is different from the usual one where one requires the renormalized $Z,\gamma$ two-point function to vanish 
at zero momentum. This condition cannot be implemented without electromagnetism, obviously. However the different result
for $\delta g$ is of no consequence in the on-shell scheme.

\subsection{Cross checks and comparison with previous results}
All these counterterms above have the correct SM limit. When $a=b=d_3=d_4=1$, all the parameters that are not present
in the SM vanish, and we are left with $\delta v^2_{div},\, \delta \lambda_{div},\,\delta \lambda_3$ and $\delta \lambda_4$.
In the SM limit, $\delta \lambda, \delta \lambda_3$ and $\delta \lambda_4$ have been checked to be exactly equal,
as it should since they all derive from the unique SM Higgs potential coupling $\lambda$ present in the
tadpole, triple and quartic self-couplings. In particular
\begin{equation}
\delta \lambda_{div,SM}=\delta \lambda_{3,SM}=\delta \lambda_{4,SM}=\frac{\Delta}{16\pi^2v^4}3(3M_W^4-3M_W^2M_h^2+M_h^4)\,.
\end{equation}
As explained before, it can be seen just by direct comparison with (\ref{eq: counter_masses_Zs}),
that the relation (\ref{eq: relation_Z_v}) is satisfied.\myspace
We can also compare our counterterms with the results previously reported in the literature. As mentioned before, 
the authors in Ref \cite{Espriu:2013fia} made
a complete study of the elastic $\omega\omega$ scattering at one-loop level, allowing only longitudinal modes 
in the internal lines.
That is, they set $g=0$ for the whole process and therefore they set the value $M_W=0$ for the vector boson mass. Our results
(\ref{eq: counter_masses_Zs}) and (\ref{eq: counter_coupling}) have been checked with those relevant for the process 
in \cite{Espriu:2013fia}
in the limit $M_W=0$.\myspace
A cruder approximation was taken in \cite{Delgado:2013hxa} where they studied all the processes including the
 $I=0$ final states but, besides setting $g=0$ and neglecting physical vector bosons in the loops, the authors took 
the limit $M_h=0$. In this limit where the self-interactions of the Higgs are absent, there is no need for 
redefinitions of $a,b$ and $v$ to absorb the one-loop divergences and
we are left with $a_4,a_5,\gamma,\delta,\eta$. Our results agree with theirs in the limit $M_W=M_h=0$, so the inclusion of the transverse gauge modes does not modify these counterterms.\myspace
We also find agreement with the results of \cite{Gavela:2014uta}, where the authors carried out the renormalization of the off-shell 
Green functions of the three processes studied in this work for the purely scalar sector of the custodial 
preserving HEFT with a light Higgs in the limit $g=0, g^{\prime}=0$ (i.e. $M_W=0$). For the comparison with our 
on-shell calculation, we have made use of the equations of motion for the Higgs and the Goldstone fields, 
omitting the leptonic contribution
\begin{equation}\label{eq: eoms}
\Box\omega=-\frac{2a}{v}\partial_{\mu}\omega\partial^{\mu}h+ \cdots\quad ,\quad \Box h=-V^{\prime}(h)+\left(\frac{a}{v}+\frac{b}{v^2}h\right)\partial_{\mu}\omega\partial^{\mu}\omega+\cdots
\end{equation}
leading to the following redefinitions of the electroweak and chiral parameters.
\begin{equation}\label{eq: redefinitions_belen}
\begin{split}
&M_h^2=\widetilde{M}_h^2-2c_{\Box\,H}\frac{\widetilde{M}_h^4}{v^2},\quad a=a_C+\frac{2\widetilde{M}_h^2}{v^2}\left(c_7-a_C\,c_{\Box\,H}\right),\\
&b=b_C+2\mu_3^v\left(c_7-a_C\,c_{\Box H}\right)+\frac{\widetilde{M}_h^2}{v^2}\left(8\,a_7-8a_C\,a_{\Box H}-4b_C\,c_{\Box H}+a_C\,c_{\Delta H}\right), \\
&\lambda_3=\frac{\mu_3^v}{3!}-\frac{\widetilde{M}_h^2}{v^2}\mu_3^vc_{\Box H}+\frac{\widetilde{M}_h^4}{v^4}\left(\frac{1}{2}c_{\Delta H}-2a_{\Box H}\right),\\
&\lambda_4=\frac{\widetilde{\lambda}}{3!}-\left(\mu_3^v\right)^2c_{\Box H}+\frac{\widetilde{M}_h^2}{v^2}\left(\mu_3^v(\frac{5}{3}c_{\Delta H}-8a_{\Box H})-\frac{4}{3}\widetilde{\lambda}c_{\Box H}\right)+\frac{4\widetilde{M}_h^4}{v^4}\left(\frac{2}{3}a_{\Delta H}-b_{\Box H}\right),\\
&a_4=c_{11},\quad a_5=c_6-\frac{a_C}{2}c_7+\frac{a_C^2}{4}c_{\Box\,H}, \quad \delta=-c_{20}+\frac{1}{2}a_Cc_{\Delta H}, \\
&\eta=-c_8+2a_Cc_{10}-4a_C^2c_9, \quad \gamma=c_{DH} \\
\end{split}
\end{equation}
where $\mu_3^v\equiv \frac{\mu_3}{v}$ and all the quantities of the form $\widetilde{X}$ represent the parameters
from their Lagrangian 
that have a direct counterpart in ours.\myspace
It is worth commenting on the relevance of off-shell calculations. First we see at once that the number of parameters
simply explodes; trying
to do phenomenology is in practice nearly impossible. Secondly, there is a large arbitrariness in using totally or partially
the equations of motion so some degree of arbitrariness is unavoidable. Finally, we have to remember that off-shell
couplings in an effective theory are devoid of any physical meaning. A glance at the l.h.s. and the r.h.s. of the
previous equivalences should suffice to convince oneself that this is not the way to go.\myspace
We have also compared our results with the more recent study \cite{Herrero:2021iqt} where the authors performed a
full (off-shell) renormalization of the one-loop Green functions involved in $W Z$ scattering (all polarizations considered)
including custodially nonpreserving operators too. When we restrict our set of counterterms to those
relevant for $W_L W_L$ scattering and take into account that custodially nonpreserving contributions are omitted,
we find our results compatible with those of \cite{Herrero:2021iqt} with two differences originating from the
inclusion by these authors of two $\mathcal{O}(p^4)$ operators,
\begin{equation}
  a_{\Box\Box}\frac{\Box h\Box h}{v^2}, \qquad a_{\Box VV}\frac{\Box h}{v}\text{Tr}\left(V_{\mu}V^{\mu}\right).
\end{equation}
These operators can be reduced by using the equation of motion of the Higgs field  at leading order in Eq. (\ref{eq: eoms}),
and are for our purposes redundant. The first of these two operators enters directly in the renormalization of the propagator
of the Higgs so, even with the same OS renormalization condition as ours, $\delta Z_h$ and $\delta M_h^2$ differ 
in a consistent way. After using the e.o.m., both operators are actually redundant and change the coefficients 
$M_h$,$\,a$ and $a_5$ following Eq. (\ref{eq: redefinitions_belen}) with $a_{\Box\Box}=c_{\Box H}$ and $a_{\Box VV}=c_7$. 
As mentioned before, from this reference we are able to compare only those counterterms participating in our elastic $W_LW_L$ scattering within 
the custodial limit ($g^{\prime}=0,\, M_Z=M_W$) and in the Landau gauge ($\xi=0$): $v^2, M_h^2,M_W^2,a,\,a_3,a_4,a_5$ and $\zeta$ (called $a_{d2}$ in their notation).\myspace
$\delta T_{div}$ is also compatible up to a different sign in the renormalization condition for the Higgs tadpole. 
We do not find agreement, though, in the counterterm associated to the $SU(2)_L$ coupling $g$, coming from the fact that in our case it is a 
derived quantity as shown in (\ref{eq: counter_g}).\myspace
One can check easily that all our additional or `anomalous' counterterms do vanish in the SM limit, while this is not in general the case for the off-shell calculations in \cite{Gavela:2014uta} and \cite{Herrero:2021iqt}.\myspace
The authors in \cite{Buchalla:2013rka} also obtained the divergences of the HEFT local operators with the
heat kernel formalism for the path integral. To this purpose, they needed to make redefinitions of the
quantum fields, in particular the Higgs field, so they were present in the canonical normalization.
These redefinitions alter the UV divergences of some operators with respect to those in our Lagrangian.
All the $\mathcal{O}(p^4)$ divergences, not affected by field renormalizations, have been checked to
coincide after some reparametrization of the chiral couplings.


\subsection{Imaginary part: The optical theorem}
The imaginary part of the NLO amplitude is obtained exactly using the optical theorem. The fact that some states can
go on shell in the process forces the presence of a physical cut in the analytic structure of an amplitude that
depends on the variable $s$ promoted to a complex quantity. This amplitude is obtained after the analytical
continuation to the whole complex plane of the Feynman amplitude depending on the centre of mass energy,
a real variable. \myspace
Given a physical amplitude $\mathcal{A}(s)$, once we know the discontinuity of the complex amplitude across the physical
cut with the usual Cutkosky rules, we find
\begin{equation}\label{eq: OT}
Im\,\mathcal{A}(s)=\sigma (s)|\mathcal{A}(s)|^2
\end{equation}
where $\sigma (s)=\sqrt{1-\frac{(M_1+M_2)^2}{s}}$ is the two-body phase space. This allows us
to compute the imaginary part of any amplitude at the one-loop level from the tree-level result.\myspace
As an example, if we are interested in computing the full $I=1$ isospin amplitude in the process
$W_L^+W_L^-\to Z_LZ_L$
\begin{equation}\label{eq: full_amp}
\mathcal{A}(W_L^+W_L^-\to Z_LZ_L)=\mathcal{A}_{tree}^{(2)}+\mathcal{A}_{tree}^{(4)}+\mathcal{A}_{loop}^{(4)}
\end{equation}
where $\mathcal{A}^{(2)}_{tree}+\mathcal{A}_{tree}^{(4)}$ is the amplitude (\ref{eq: tree_WW}) and $\mathcal{A}_{loop}^{(4)}$ 
is the full one-loop amplitude
\begin{equation}
\mathcal{A}^{(4)}_{loop}=Re \left[\mathcal{A}^{(4)}_{loop}(\omega^+\omega^-\to zz)\right]+i \sigma (s) |\mathcal{A}_{tree}^{(2)}|^2 .
\end{equation}
This procedure is not necessary but speeds up the calculation.

\section{Unitarization}\label{sec:unitarization}
Departures from the SM such as those described by the HEFT unavoidably result in a loss of
unitarity. Amplitudes typically exhibit a bad ultraviolet behavior leading to cross sections
that grow too fast with the energy $\sqrt s$ and quickly violate the unitarity bounds.
While this is well known, it is sometimes forgotten that this fast growth of the amplitudes
results in a hypersensitivity to deviations of the coefficients of the HEFT with respect to
their SM values. We analyze this in some more detail in the next subsections. \myspace
Once we have built the fixed isospin amplitudes using Eq. (\ref{eq: Ts}), we can obtain the amplitude
with $J$ total angular momentum with the corresponding partial wave
\begin{equation}\label{eq: projection}
t_{IJ}(s)= \frac{1}{32K\pi}\int_{-1}^{1}d(\cos\theta)P_J(\cos\theta)T_I(s,\cos\theta)
\end{equation}
using the center of mass relations $t=-(s-4M_W^2)(1-\cos\theta)/2$ and $u=-(s-4M_W^2)(1+\cos\theta)/2$. $K$ is a constant whose value is $K=2$ or $1$
depending on whether the particles participating in the process are identical or not.\myspace
The way we compute the fixed isospin amplitudes using Feynman diagrams from the Lagrangian (\ref{eq: lag_complete}),
leads to a perturbative expansion of the form
\begin{equation}\label{eq: texpansion}
t_{IJ}(s)\approx t^{(2)}_{IJ}+t^{(4)}_{IJ}+\ldots
\end{equation}
which for the $I,J=1,1$ case satisfies perturbatively the optical theorem (\ref{eq: OT}) 
\begin{equation}\label{eq: OT_pert}
\begin{split}
Im\left(t^{(2)}_{11}\right)&=0\\
Im\left(t^{(4)}_{11}\right)&=\sqrt{1-\frac{4M_W^2}{s}} |t^{(2)}_{11}|^2
\end{split}
\end{equation}

\subsection{Amplitudes at high energies}

\begin{figure}
\centering
\includegraphics[clip,width=12cm,height=9cm]{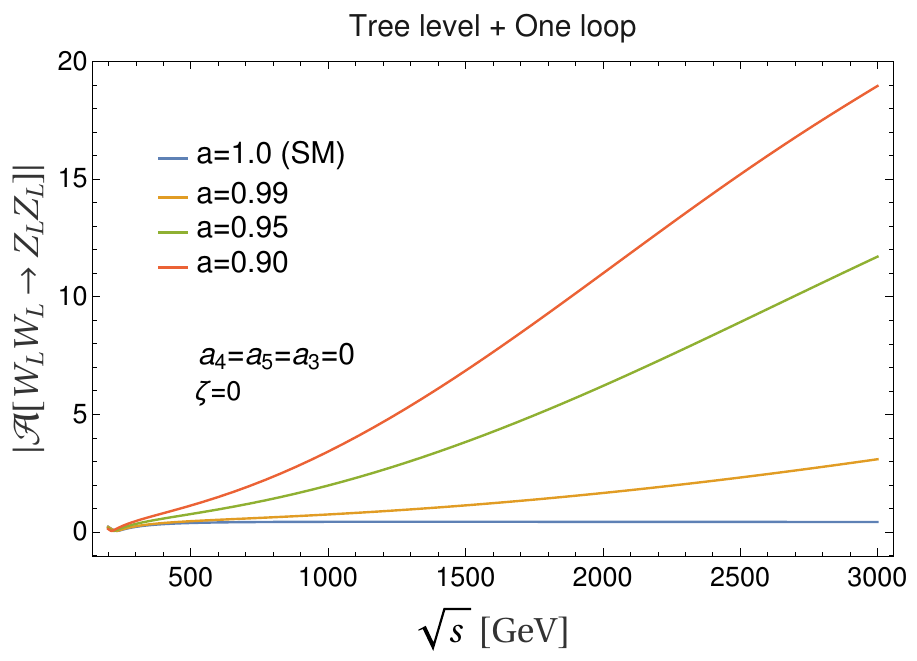}
\caption{\small{Plot of the modulus of the elastic vector boson scattering (VBS) amplitude in longitudinal 
polarization ($W_LW_L\to Z_L Z_L$) versus the center of mass energy $\sqrt{s}$ for some values of the chiral parameter $a$ at a fixed scattering angle $\cos\theta=0.3$. 
It can be seen how small departures for the SM value ($a=1$) leads to a quick violation of unitarity within 
the HEFT regime of validity. All the $\mathcal{O}(p^4)$ couplings contributing to the process 
($a_3,\,a_4,\, a_5,\,\zeta$) are set to zero}}
\label{fig: real_WWWW_a}
\end{figure}

\begin{figure}
\centering
\includegraphics[clip,width=8cm,height=6cm]{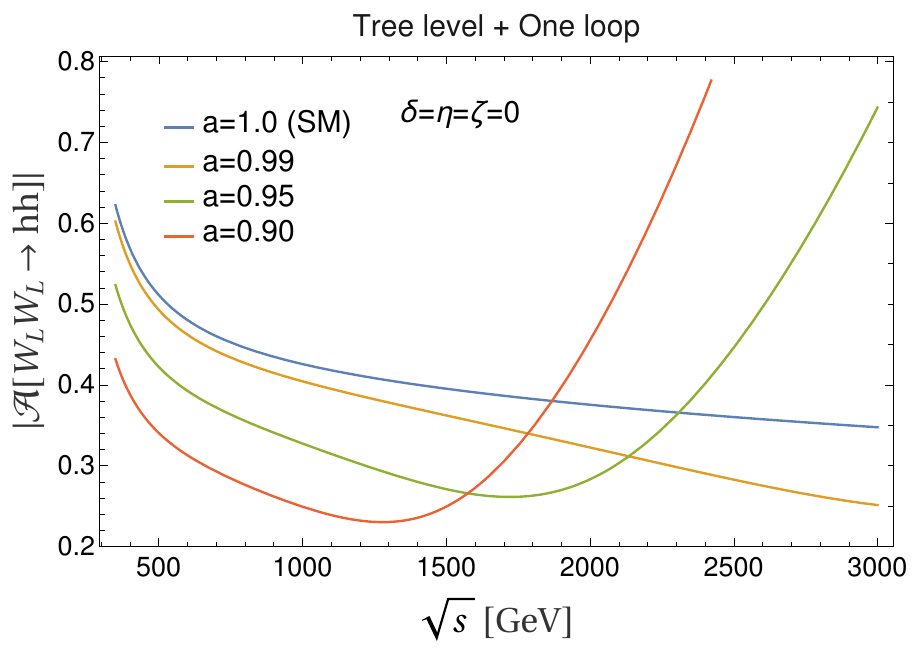} %
\includegraphics[clip,width=8cm,height=6cm]{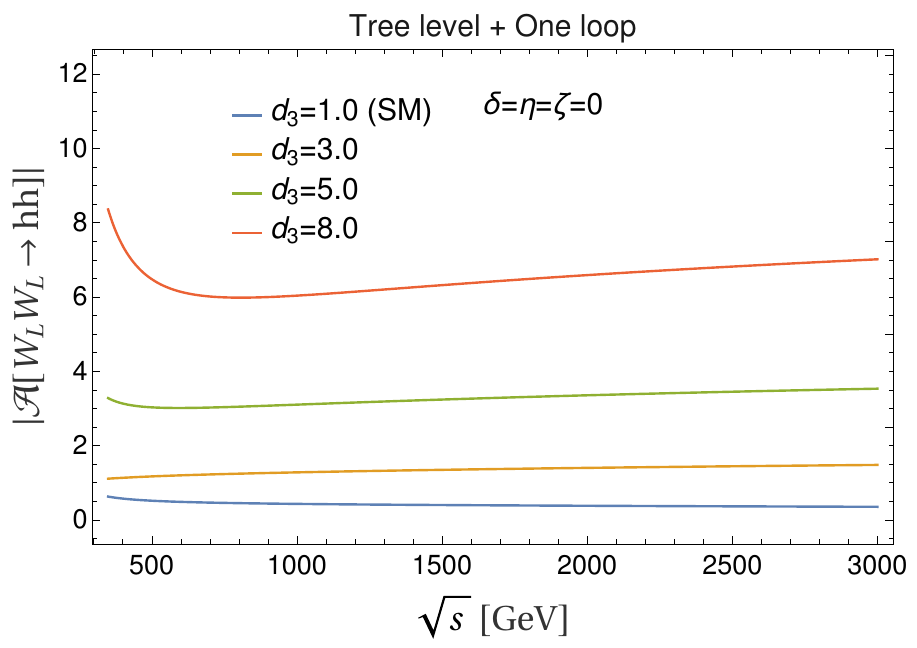} %
\caption{\small{Plot of the modulus of the $W_LW_L\to hh$ amplitude versus the center of mass energy $\sqrt{s}$ 
for some values of the chiral parameter $a$ (left) and the trilinear Higgs coupling $\lambda_3=d_3\lambda$ (right) at a 
fixed scattering angle $\cos\theta=0.3$. It can be seen how departures from the SM limit do not lead to an obvious 
bad UV behavior of the amplitude in the second case. Although it could seem that in the left panel there is no bad high energy limit for the case $a=0.99$, the reality is that the amplitude acquires an 
unphysical behaviour for a scale just above the cut-off of the theory (around $4$ TeV), a region not shown in the plot. 
In each figure, the remaining parameters do not vary and are set to the corresponding SM values.}}%
\label{fig: real_WWHH}%
\end{figure}

\begin{figure}
\centering
\includegraphics[clip,width=12cm,height=8.5cm]{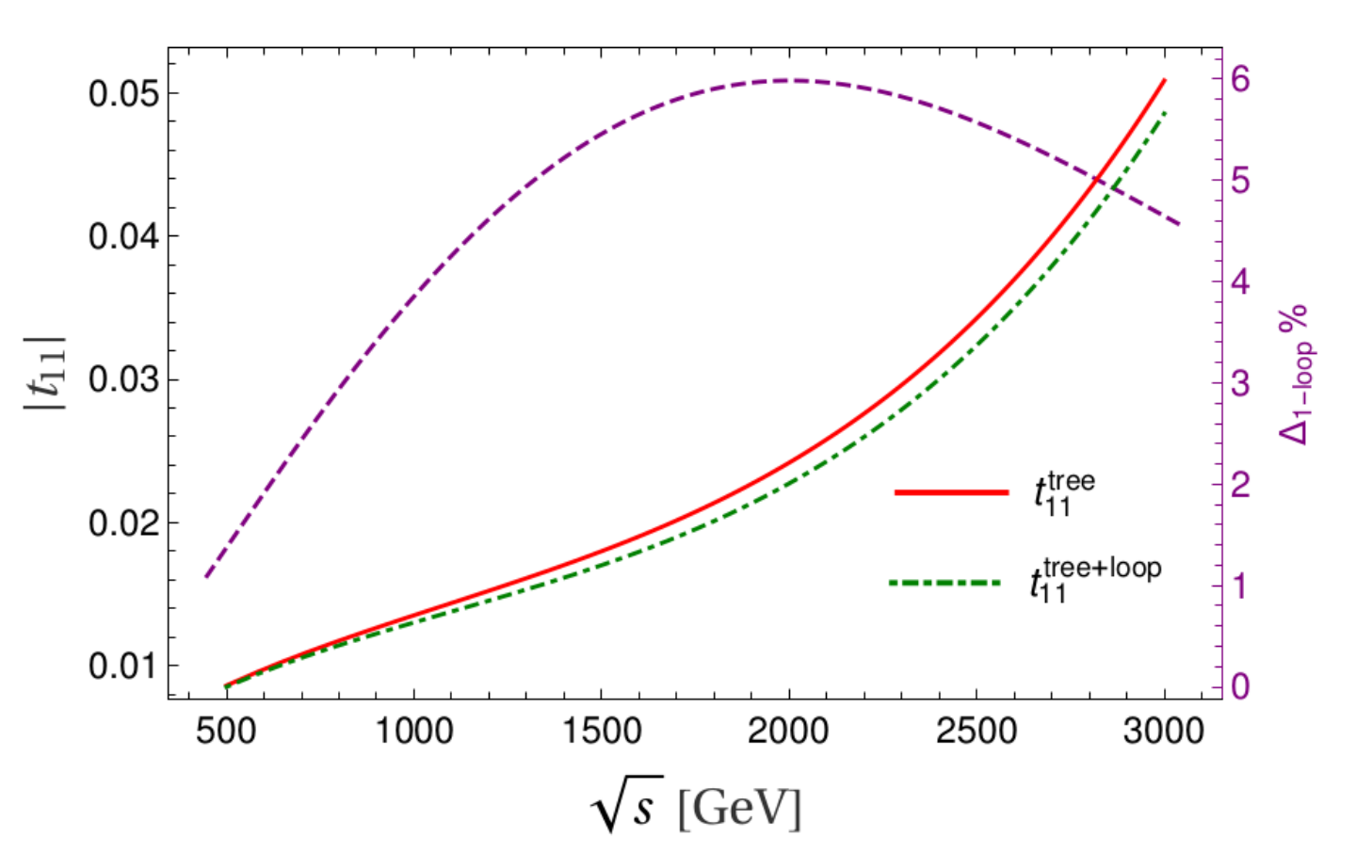}
\caption{\small{(Left axis) Plot of the modulus of the vector-isovector partial wave at tree level (solid red line) and tree + one 
loop level (dot-dashed green line) versus the center of mass energy $\sqrt{s}$. (Right axis) Plot of the percentage represented 
by the 1-loop contribution (purple dashed line), $\Delta_{1-loop}$ in Eq. (\ref{eq: treevsloop}) versus the center of 
mass energy $\sqrt{s}$ in absolute value. The curves are depicted for $a_4=10^{-4}$ and the rest of the parameters are set to their SM values.}}
\label{fig: t11_loopvstree}
\end{figure}

In Fig. \ref{fig: real_WWWW_a} we plot the modulus computed at the tree plus one-loop level for the process
$W_L W_L \to Z_L Z_L$ for various values of the parameter $a$ at a fixed scattering angle $\cos\theta=0.3$. Departures from the SM value $a=1$ result 
in a clear bad high-energy behavior.  
This same behavior is seen in the remaining $2\to 2$ processes. For instance the modulus of the amplitude for the process $W_L W_L \to hh$ is depicted in the left panel of Fig. \ref{fig: real_WWHH} for the same values
of the parameter $a$ parameterizing the Higgs-vector boson coupling in the HEFT.
\myspace
On the contrary, this same modulus of the $W_L W_L \to hh$ amplitude (Fig. \ref{fig: real_WWHH}) shows 
a milder dependence on the parameter $\lambda_3$ of the Higgs potential. For the SM value $a=1$, modifying $\lambda_3$
does not show obvious signs of bad high-energy behavior. At this level, this derives from the fact that this coupling is 
momentum independent. Note that this coupling is very poorly constrained so the overall uncertainty of
the amplitude is accordingly large. \myspace
Higher-loop calculations will only worsen the high-energy behavior. It is thus clear that, except for tiny
deviations from the SM, as soon as one enters the multi TeV region, the perturbative treatment
is unreliable. Therefore, checking for constraints on the anomalous couplings present
in the HEFT by just looking at growing cross sections is risky but may be justified (if the deviations are small) or 
plain wrong (if the anomalous coupling constants deviate significantly from their SM values).
It is clear that physical amplitudes --even beyond the SM-- are necessarily unitary, meaning that
in the HEFT higher-loops contributions have to be somehow summed up to render a reasonable
high-energy behavior, which of course will be different from the SM one, but still in accordance with
the general principles of field theory.
We conclude that unitarization is necessary to compare the predictions of the
HEFT with those of the SM \textit{vis-\`a-vis} the experiments at very high energies, particularly when we are close to the HEFT UV cutoff.\myspace
The bad-energy behavior can also be seen at the partial wave level before unitarization. The modulus of the vector-isovector contribution up to the NLO in the expansion (\ref{eq: texpansion}) is shown in Fig. \ref{fig: t11_loopvstree} 
for a small departure from the SM values via the parameter $a_4=10^{-4}$ (the rest of the parameters are set to their 
corresponding SM values). In that same figure (in the right axis), the contribution in percentage (in absolute value) 
to the one-loop partial wave is shown with respect to the full tree level [$\mathcal{O}(p^2)$+$\mathcal{O}(p^4)$] with 
the following definition
\begin{equation}\label{eq: treevsloop}
\Delta_{1-loop}=100 \cdot \Bigg | \frac{|t_{11}^{tree+loop}|-|t^{tree}_{11}|}{|t_{11}^{tree}|}\Bigg |
\end{equation}
\noindent
The contribution of the one-loop level to the full partial wave turns out to be negative as it can be seen in the figure 
(the green dot-dashed line representing the full amplitude is always below the tree level contribution in solid red), reaching 
a maximum value  $\sim 6\%$ around the $2$ TeV region. Here the tree level contribution includes, as mentioned, 
both the $O(p^2)$ and the $O(p^4)$ pieces. As was already notice
in previous works \cite{Espriu:2013fia}, as soon as one departs 
from the SM the later quickly dominate the real part of the $O(p^4)$ contribution 
(yet another reason why using the ET is justified).

\subsection{ The inverse amplitude method (IAM) }
\begin{figure}
\centering
\includegraphics[clip,width=16cm,height=8.5cm]{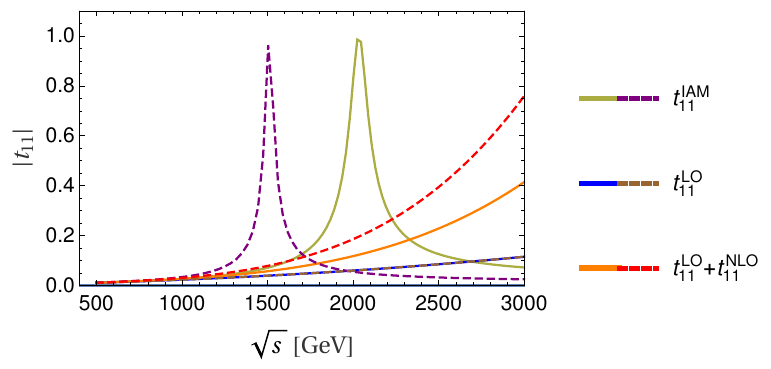}
\caption{\small{Plot of the IAM and partial wave amplitudes of the vector-isovector channel $I,J=1,1$ for the two benchmark points $BP1^{\prime}$(dashed) and $BP2^{\prime}$(solid) defined in Table \ref{table: benchmark_points} as a function of the centre of mass energy $\sqrt{s}$. The notation of LO corresponds to the chiral order two $\left(t_{11}^{(2)}\right)$ and the NLO to the chiral order four $\left(t_{11}^{(4)}\right)$ partial waves in Eq. (\ref{eq: texpansion}). Since the two benchmark points represented here share the value of $a$, the two LO lines, independent of the $\mathcal{O}(p^4)$ parameters, coincide.}}
\label{fig: IAM_PW}
\end{figure}

The expansion in terms of the external momentum typically leads very quickly to a violation of unitarity and,
in order to make realistic predictions, unitarization techniques need to be used. In our case we choose
to make use of the IAM~\cite{Truong:1988zp, Dobado:1989qm, Dobado:1996ps, Oller:1997ng, Guerrero:1998ei, Oller:1998hw, Dobado:2001rv, Corbett:2015lfa, Garcia-Garcia:2019oig} 
in order to unitarize the
partial waves (and, eventually, the amplitudes). The IAM is really successful in
predicting the features of the rho meson resonance studying low-energy QCD with pion-pion scattering
and it has also been extensively used in HEFT analysis.\myspace
The method consists in building the following IAM amplitude up to NLO
\begin{equation}\label{eq: IAM}
t^{IAM}_{IJ}\simeq\frac{\left(t^{(2)}_{IJ}\right)^2}{t^{(2)}_{IJ}-t_{IJ}^{(4)}}
\end{equation}
which perturbatively satisfies (\ref{eq: texpansion}) and the desired unitarity condition
$t_{IJ}(s) = \frac{i}{2}( 1 - \eta(s) e^{2i\delta(s)})$, where  $0<\eta<1$ is the ineslasticity.\myspace
A pole in the unitary amplitude (\ref{eq: IAM}) appears when, for some complex value of $s_R$ (\ref{eq: resonance_condition}), 
\begin{equation}\label{eq: resonance_condition}
t_{IJ}^{(2)}(s_R)-t_{IJ}^{(4)}(s_R)=0.
\end{equation}
This pole, if present, is interpreted as a resonance with quantum numbers $I,J$ and features $M_R$ and $\Gamma_R$, 
these lasts given \textit{a la} Wigner by the position of the pole in the complex plane $s_R=\left(M_R\ih \Gamma_R\right)^2$.
We will consider for this study and future ones only the lowest partial wave in $I=0,1$ channels and refer to these
resonances as scalar-isoscalar for the poles in $t^{IAM}_{00}$ and vector-isovector for $t^{IAM}_{11}$.\myspace
One nicety of the IAM, besides assuring unitarity, is that the poles can be interpreted as dynamically generated
resonances appearing after the resummation of infinite bubbles chain $WW\to ZZ\to WW\to\ldots \to ZZ$ (in the $I=1$ channel)
as it can be understood diagrammatically from the perturbative expansion of (\ref{eq: IAM}).\myspace
An example of the recovery of unitarized amplitudes by the IAM is depicted in Fig. \ref{fig: IAM_PW}. In that illustration we show both IAM and partial wave amplitudes for the two benchmark points defined in the next section Table \ref{table: benchmark_points}, $BP1^{\prime}$ and $BP2^{\prime}$. It can be seen how the inclusion of the NLO contribution leads to an even quicker violation of the unitarity of the partial wave in the UV regime of the theory. This unphysical high-energy behavior is tamed by means of the IAM amplitude defined in Eq. (\ref{eq: IAM}), exhibiting resonances for a BSM model. Fig. \ref{fig: IAM_PW} also shows the importance of the next to leading order versus the leading order contribution, reaching a $40\%-60\%$ relative size difference near the cutoff of the theory.\myspace
In the $I=0$ channel, two Higgs intermediate states are possible and for that one needs the machinery of 
coupled channels. \myspace 
The IAM can be extended to the coupled channel case too (see 
\cite{GomezNicola:2001as, Oller:1997ng}), particularly if all the different channels have the same thresholds.  
From the perturbative expansion
\begin{equation}\label{eq: T_pert}
T_{IJ}=T_{IJ}^{(2)}+T_{IJ}^{(4)}+\dots \\ 
\end{equation}
a natural generalization of the IAM method gives
\begin{equation}\label{IAMforF}
 T_{IJ}^{\rm IAM}=T_{IJ}^{(2)}(T_{IJ}^{(2)}-T_{IJ}^{(4)})^{-1}T_{IJ}^{(2)}
\end{equation}
which satisfies exact multichannel elastic unitarity on the right cut
\begin{equation} \label{multichannelUnit}
Im \, T_{IJ}^{\rm IAM} = T_{IJ}^{\rm IAM}(T_{IJ}^{\rm IAM})^\dagger .
\end{equation}
The IAM has been extensively used to describe low-energy meson-meson scattering where it has proven to be extremely 
successful. With a very small set of parameters, it is able to  describe many different channels including their first 
resonances \cite{Delgado:2015kxa, Garcia-Garcia:2019oig, GomezNicola:2001as, Oller:1997ng}. In the case of coupled channels, 
the different amplitude matrix
elements (partial waves) $(T_{IJ})_{ij}(s)$ correspond 
to different reactions having the same quantum numbers $IJ$. Clearly, if there is a resonance in one of the channels
it should appear also in all the others since physically these resonances can be produced in any of the reactions.\myspace
While for single-channel unitarization the IAM is well grounded and relies on a minimal set of assumptions
(see e.g. \cite{Delgado:2015kxa, GomezNicola:2001as, Oller:1997ng, Salas-Bernardez:2020hua}), there is no really unambiguous way of 
applying the IAM to the case where there are coupled channels
with different thresholds. We shall adhere to the simplest choice that
consists in assuming the previous expressions to remain valid also in the present analysis.
This can be justified heuristically on
the grounds that $M_W$ is not too different from $M_h$. This is again a good justification of
the need to include all polarizations of the vector boson with a mass $M_W$ in the calculation. \myspace
In addition, it should be stated that  the decoupling of the two $I=0$ channels in the case $a^2=b$ taking
place when the equivalence theorem is used and physical $W_L$ are replaced by the corresponding
Goldstone bosons does not hold in the exact calculation. \myspace
The results for the $IJ=00$ channel will be reported in a separate publication. Here we will
concentrate in the modifications that the inclusion of the transverse mode propagation
of the vector bosons with a mass $M_W$ and the appearance of new effective couplings in the HEFT induce
in the $IJ=11$ channel.

\subsection{Vector resonances}
In order to see the relevance of including the propagation of transverse modes, we focus  on vector resonances
with quantum numbers $I,J=1,1$ in VBS. We shall compare the new results with those obtained previously.\myspace
From Eq. (\ref{eq: isos_I1}) and (\ref{eq: Ts}), the fixed isospin amplitudes in the chiral expansion, $T_1^{(2)}$
and $T_1^{(4)}$ are obtained. $T_1^{(2)}$ using $\mathcal{A}_{tree}^{(2)}(p_1,p_2,p_3,p_4)$ and $T_1^{(4)}$ with
$\mathcal{A}_{tree}^{(4)}(p_1,p_2,p_3,p_4)+Re \left[\mathcal{A}_{loop}(\omega^+\omega^-\to zz)\right](p_1,p_2,p_3,p_4)$.
Using Eq. (\ref{eq: projection}) and (\ref{eq: OT_pert}) perturbatively, we find the partial wave  for $I,J=1,1$
\begin{equation}\label{eq_ my_t11}
\begin{split}
t^{(2)}_{11}=&\frac{1}{64\pi}\int_{-1}^{1}d(\cos\theta)\cos\theta\,T^{(2)}_1(s,\cos\theta)\\
Re\left[t^{(4)}_{11}\right]=&\frac{1}{64\pi}\int_{-1}^{1}d(\cos\theta)\cos\theta\,T^{(4)}_1(s,\cos\theta)\\
Im\left[t_{11}^{(4)}\right]=&\sqrt{1-\frac{4M_W^2}{s}}|t_{11}^{(2)}|^2\\
\end{split}
\end{equation}
where the Legendre polynomial $P_1(\cos\theta)=\cos\theta$ has been used.\myspace
The vector-isovector resonances, if present, are located by searching for poles of the unitary IAM amplitude (\ref{eq: IAM})
i.e. looking for solutions of Eq. (\ref{eq: resonance_condition}).
\myspace
Let us first of all investigate how the proper inclusion of the transverse modes (i.e. $g\neq 0$) influence the results obtained in
the extreme ET limit. Below we provide results for $g=0$ and $g=2M_W/v$. The benchmark points correspond to those
used in \cite{Delgado:2017cls}.
\begin{table}[h!]
\begin{tabular}{|c|c|c|c|c|c|}
\hline
    $\sqrt{s_V} \, (GeV)$ & $\quad g=0\quad $  &  $\quad g\neq 0\quad$   & $ \quad a \quad $   & $a_4\cdot 10^4$ & $a_5\cdot 10^4$ \\ \hline
BP1  & $\quad 1476\ih 14 \quad$ & $\quad 1503\ih 13 \quad$       & 1   &   3.5          & -3   \\ \hline
BP2  & $2039\ih 21$ & $2087\ih 20$        & 1   &   1            & -1   \\ \hline
BP3  & $2473\ih 27$ & $2540\ih 27$        & 1   &   0.5          & -0.5 \\ \hline 
BP1$^{\prime}$ & $1479\ih 42 $ & $1505\ih 44$        & 0.9 & 9.5            & -6.5  \\ \hline
BP2$^{\prime}$ & $1981\ih 97$ & $2025\ih 98$        & 0.9 & 5.5            & -2.5   \\ \hline
BP3$^{\prime}$ & $2481\ih 183$ & $2547\ih 183$        & 0.9 & 4              &  -1  \\ \hline
\end{tabular}
\caption{{\small
  Values for the location of the vector poles $\sqrt{s_V}=M_V-\frac{i}{2}\Gamma_V$ found in all the benchmark points 
of reference \cite{Delgado:2017cls} once the transverse modes are included $(g\neq 0)$. }}\label{table: benchmark_points}
\end{table}
As it can be seen {\em ceteris paribus} the inclusion of the gauge boson masses systematically 
increases the masses of the resonances
by a few per cent. The modifications in the widths are not significant.
In these calculations $b=a^2$, and both $a_3$ and $\zeta$ have been set to zero.

\subsection{Checking unitarity}
\begin{figure}
\includegraphics[clip,width=0.6\textwidth]{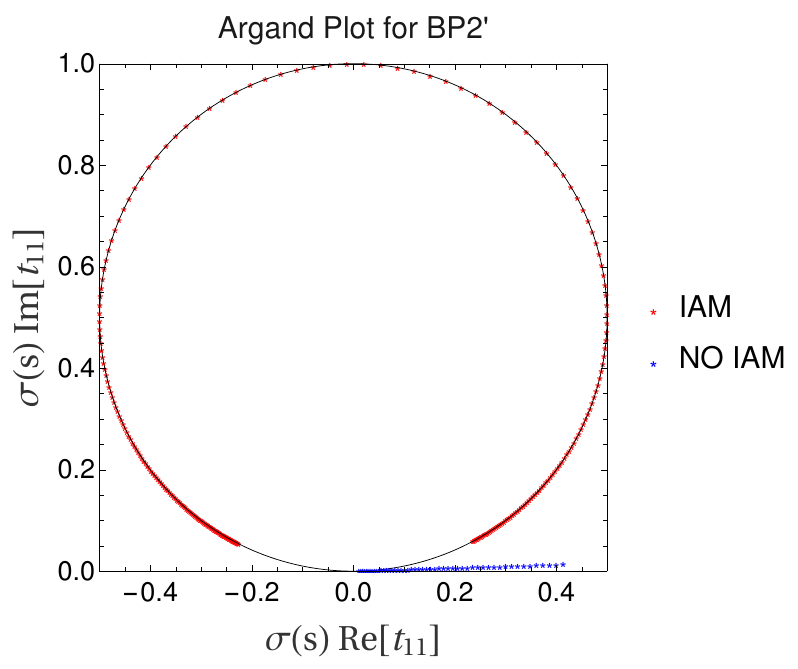}
\caption{\small{Argand plot showing the unitary VBS amplitude (red points) for the values of BP2$^{\prime}$ from 
Table \ref{table: benchmark_points}. Due to the elasticity of the process, the IAM amplitude lies exactly on 
the unitarity limit, i.e, the circumference of radius $1/2$ centered at $(0,1/2)$. The amplitude before 
applying the IAM is also present (blue points) and obviously lies entirely outside the unitarity condition.}}
\label{fig: argand}
\end{figure}

As a check of the good unitarity behavior of the amplitudes obtained in the IAM and the validity of the approximations made we plot
the partial wave for complex values of the kinematical variable $s$ in the $IJ=11$ channel.
There are no threshold in this channel beyond the elastic channel and the results must lie
accordingly in a circumference of radius 1/2 centered at $s= i/2$.
This is shown in Fig. \ref{fig: argand}. We also plot the results obtained for the 
same $IJ=11$ channel in perturbation theory without resummation.  They obviously
violate the unitarity bound. The plot correspond to the values of the benchmark point BP2$^{\prime}$ from 
Table \ref{table: benchmark_points}, corresponding to $a=0.9,\,a_4=5.5\cdot10^{-4}$ and $a_5=-2.5\cdot 10^{-4}$, 
within the recent CMS experimental bounds of Table \ref{table: exp}.

\subsection{Influence of the new HEFT constants}
The inclusion of transverse modes in the calculation of the dispersive real part of the amplitude
leads unavoidably to consider the case where $M\neq 0$. We have seen how this changes to some extent
the location and widths of the amplitudes.  On the other hand, the inclusion of the transverse modes
leads to the appearance of new counterterms in the HEFT. In the channel $IJ=11$ two new low-energy
(`anomalous') couplings appear; $a_3$ and $\zeta$.  Let us see how their presence may affect
the previous results.\myspace
For this analysis we focus in the four lightest resonances of Table \ref{table: benchmark_points}. 
This is BP1, BP2, BP1$^{\prime}$ and BP2$^{\prime}$
\begin{table}[h!]
\begin{tabular}{|c|c|c|c|c|c|}
\hline
    $\sqrt{s_V} \, (GeV)$&$a_3=0$ & $a_3=0.1$  &  $a_3=-0.1$   & $ a_3=0.01 $   & $a_3=-0.01$ \\ \hline
BP1 & $\quad 1503\ih 13 \quad$ & $\quad 1795\ih 11 \quad$ & $\quad 1215\ih 15 \quad$        &  $ \quad 1532\ih 13 \quad$  &   $\quad 1474\ih 13 \quad$             \\ \hline
BP2 & $2087\ih 20$ & $2721\ih 15$ &   $1505\ih 23$      &   $2150\ih 19$ &  $2025\ih 21$              \\ \hline 
BP1$^{\prime}$& $1505\ih 44$& $1663\ih 46$ &  $1335\ih 43$       & $1520\ih 44$ &     $1488\ih 44$   \\ \hline
BP2$^{\prime}$&$2025\ih 98$ & $2278\ih 104$ &   $1752\ih 89$      & $2052\ih 98$ &      $1999\ih 97$  \\ \hline

\end{tabular}
\caption{{\small
   Values for the location of the vector poles $\sqrt{s_V}=M_V-\frac{i}{2}\Gamma_V$ found in all the benchmark 
points of reference \cite{Delgado:2017cls} 
for different values of $a_3$ and $g\neq 0$. The chiral parameter $\zeta$ is set to zero.  }}\label{table: benchmark_points_a3}
\end{table}
\begin{table}[h!]
\begin{tabular}{|c|c|c|c|c|c|}
\hline
    $\sqrt{s_V} \, (GeV)$ &$\zeta=0$ & $\zeta=0.1$  &  $\zeta=-0.1$   & $\zeta=0.01 $   & $\zeta=-0.01$ \\ \hline
BP1  & $\quad 1503\ih 13 \quad$ &   $\quad 1637\ih 13 \quad$      & $\quad 1377\ih 14 \quad$   & $\quad 1516\ih 13 \quad$   &  $\quad 1489\ih 13 \quad$          \\ \hline
BP2  & $2087\ih 20$ &     $2393\ih 18$    & $1809\ih 22$   &  $2117\ih 20$   &  $2058\ih 21$         \\ \hline 
BP1$^{\prime}$ & $1505\ih 44$ &  $1570\ih 46$       & $1439\ih 43$ & $1510\ih 45$  &  $1497\ih 45$   \\ \hline
BP2$^{\prime}$ & $2025\ih 98$ &   $2136\ih 100$       & $1915\ih 94$ & $2036\ih 98$  &  $2014\ih 97$   \\ \hline

\end{tabular}
\caption{{\small
    Values for location of the vector poles $\sqrt{s_V}=M_V-\frac{i}{2}\Gamma_V$ found in all the benchmark points of reference \cite{Delgado:2017cls} for different values of $\zeta$ and $g\neq 0$. The chiral parameter $a_3$ is set to zero.  }}\label{table: benchmark_points_zeta}
\end{table}
We see from the previous results that, of the two new parameters (not previously considered in unitarization
analysis), $a_3$ is most relevant as it can be seen in Tables \ref{table: benchmark_points_a3} and \ref{table: benchmark_points_zeta}. Positive values of $a_3$ tend to increase the mass of the vector resonance and make it even 
narrower, making its detection harder. Negative values of $a_3$ work in the opposite direction. Although the bounds on
$a_3$ allow it, the value $|a_3| = 0.1$ may be too large, and we also provide $M_V$ and $\Gamma_V$ for   
$|a_3| = 0.01$. If $a_3$ happened to be of the same order as the current bounds for $a_4$ and $a_5$, its effect would be
subleading. The influence of $\zeta$ appears to be less than that of $a_3$ but the qualitative behavior remains.

\section{Conclusions}
One of the main results of this paper is the determination of the one-loop quantum corrections to 
all the relevant $2\to 2$ processes that are relevant to two-Higgs production via the scattering 
of electroweak gauge bosons in the HEFT. The calculation has been explicitly performed in 
the 't Hooft-Landau gauge, although physical amplitudes are gauge independent.\myspace
In our work, for the first time, a diagrammatic computation of all the on-shell $2\to 2$ processes relevant 
for two-Higgs production is presented. In the one-loop calculation, both transverse and longitudinal 
polarized modes are included. In
the on-shell scheme this necessarily leads to considering the physical values for the Higgs and weak gauge
boson masses\footnote{However, in order to be able to use safely exact isospin relations we work in
  the custodial limit neglecting electromagnetism, i.e., $g^\prime =0$.}. The resulting amplitudes 
are then unitarized and we analyze the characteristics of the dynamical resonances appearing. 
An interesting result is that, after unitarization of the
partial waves, the effect of including the gauge boson masses is small but significant increasing the 
mass of the vector resonances typically in the range
2 to 3 \%. The widths are unchanged.\myspace
The introduction of the transverse degrees of freedom of the gauge bosons also implies the need to consider
additional effective couplings that had not been previously considered in unitarization studies. 
In elastic $WW\to WW$ scattering, there are two new effective couplings that become relevant. While
traditionally the effective couplings $a_4$ and $a_5$ have been regarded as driving the masses of
dynamical resonances, it turns out that the coupling $a_3$ (that plays a role only if the {\em a priori}
subdominant transverse modes are included) is relevant too. It should also be mentioned that while
$a_4$ and $a_5$ are by now fairly constrained by LHC analysis, the bounds on $a_3$ are still rather loose.
We believe this makes the present study particularly relevant.\myspace
The calculation is done on shell, which is what is required for a useful experimental comparison. Keeping
redundant operators results in a proliferation of couplings of which only a handful are useful. We also see
that the most influential coefficients in the effective Lagrangian are those surviving the extreme ET limit.
There is some logic behind this, but it is reassuring to check it in a detailed calculation.\myspace
In the present paper, we have focused on the impact of the new contributions in the vector-isovector channel
and have postponed the consideration of the more involved scalar-isoscalar one to a future publication. Unitarization 
of the latter, that requires a full use of the coupled channel formalism, is most relevant in order to
be able to constraint some of the Higgs couplings.

\section*{Acknowledgments}
We thank Oscar Cat\`a and Rafael Delgado for useful discussions. 
We acknowledge financial support from the State Agency for Research of the Spanish Ministry of Science and Innovation
through the “Unit of Excellence Mar\'ia de Maeztu 2020-2023” award to the Institute of Cosmos Sciences (CEX2019-000918-M)
and  from PID2019-105614GB-C21 and  2017-SGR-929 grants.




\begin{small}

\bibliographystyle{utphys.bst}
\bibliography{bibliography}

\end{small}

\end{document}